\documentclass[twocolumn,epjc3]{svjour3}  
\smartqed  
\RequirePackage{graphicx}
\RequirePackage{amsmath}
\RequirePackage{amssymb}
\RequirePackage{xspace}
\catcode`~=11 
\newcommand{\urltilde}{\kern -.15em\lower .7ex\hbox{~}\kern .04em}
\catcode`~=13 

\providecommand*{\dd}%
{\ensuremath{\mathrm{d}}}

\newcommand{\Tnoem}[1]{\ensuremath{T_M^{#1}}}
\newcommand{\Tsud}{\ensuremath{T_M}}
\newcommand{\Amp}{\ensuremath{A}}
\newcommand{\spinkin}{\ensuremath{{\cal S}}}

\DeclareMathOperator{\Tr}{Tr\,}
\providecommand*{\eu}%
{\ensuremath{\mathrm{e}}}

\newcommand{\eg}{\emph{e.g.}}
\newcommand{\ie}{\emph{i.e.}}

\newcommand{\pytppp}{P\protect\scalebox{0.8}{YTHIA}8\xspace}

\newcommand{\particle}[1]{\ensuremath{\mathrm{#1}}}
\newcommand{\el}{\particle{e}}
\newcommand{\tee}{\ensuremath{\el^+\el^-}\xspace}

\newcommand{\fig}[1]{\ref{#1}}
\newcommand{\figref}[1]{figure~\fig{#1}}

\newcommand{\sect}[1]{\ref{#1}}
\newcommand{\sectref}[1]{section~\sect{#1}}

\newcommand{\eq}[1]{(\ref{#1})\xspace}
\newcommand{\eqsref}[1]{eqs.~\eq{#1}\xspace}

\newcommand{\appref}[1]{appendix~\sect{#1}}

\newcommand{\GeV}{\mbox{GeV}}
\newcommand{\TeV}{\mbox{TeV}}

\journalname{Eur. Phys. J. C}
\begin{document}

\title{{\hfill \normalsize \rm{MCnet-16-35}\\ \medskip \\} Generation of central exclusive final
  states\thanksref{t1}
}


\author{Leif L\"{o}nnblad\thanksref{e1,addr1}
        \and
        Radek \v{Z}leb\v{c}\'{i}k\thanksref{e2,addr2} 
}

\thankstext{t1}{   Work supported in part by the MCnetITN FP7 Marie
    Curie Initial Training Network, contract PITN-GA-2012-315877, and
    the Swedish Research Council (contracts 621-2012-2283 and
    621-2013-4287) }
\thankstext{e1}{e-mail: Leif.Lonnblad@thep.lu.se}

\thankstext{e2}{e-mail: zlebcr@mail.desy.de}


\institute{Dept.~of Theoretical Physics,
  S\"olvegatan 14A, S-223 62  Lund, Sweden \label{addr1}
           \and
           Institute of Particle and Nuclear Physics,
  V Hole\v{s}ovi\v{c}k\'{a}ch 2, Prague 8, Czech Republic \label{addr2}
}

\date{Received: date / Accepted: date}

\maketitle

\begin{abstract}
    We present a scheme for the generation of central
    exclusive final states in the \pytppp program. The implementation
    allows for the investigation of higher order corrections to 
    such exclusive processes as approximated by the initial-state
    parton shower in \pytppp. To achieve this, the spin and colour
    decomposition of the initial-state shower has been worked out, in
    order to determine the probability that a partonic state generated
    from an inclusive sub-process followed by a series of initial-state
    parton splittings can be considered as an approximation
    of an exclusive colour- and spin-singlet process.

    We use our implementation to investigate effects of parton showers
    on some examples of central exclusive processes, and find sizeable
    effects on di-jet production, while the effects on \eg\ central
    exclusive Higgs production are minor.
\end{abstract}

\section{Introduction}
\label{sec:intro}

\begin{figure}
  \centering
  \includegraphics{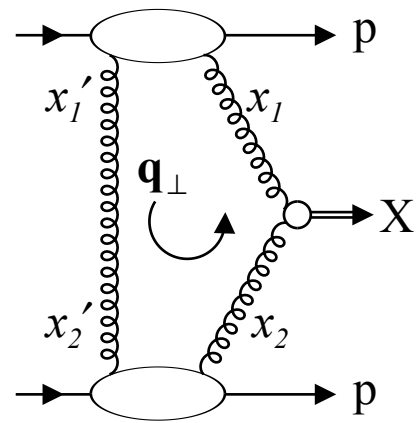}
\caption{ The basic diagram for a general
    central exclusive process $pp\to p+X+p$.}
\label{fig:exdiagram}
\end{figure}

Compared to the fairly clean environment of \tee annihilation, proton
collision events are in general very messy, especially at the LHC at
high luminosity. Even at lower luminosity where pile-up events are
absent, the existence of multiple soft interactions and initial-state
parton showers means that any hard sub-process of interest will be
obscured by soft and semi-hard hadrons smearing the
measurements. However, for some rare events, a central colour-singlet
hard sub-process may appear in complete isolation, with rapidity gaps
on both sides stretching all the way out to the (quasi-) elastically
scattered protons, giving a nice and clean environment to study its
properties.

Such \textit{Central Exclusive Processes} (CEPs) have been extensively
studied in the so-called Durham formalism, first described by Khoze,
Martin and Ryskin in \cite{Khoze:2000cy} and reviewed in detail in
\cite{Harland-Lang:2014lxa}. The simplest such process is Higgs
production, where two gluons in a colour- and spin-singlet state fuse
together via a top-quark loop into a Higgs particle as outlined in
\figref{fig:exdiagram}. With an additional virtual exchange of a
(semi-hard) gluon, the net colour exchange between the colliding
protons can be zero and we may end up with a very simple and clean
final state consisting only of two (quasi-) elastically scattered
protons along the beam pipe and the Higgs decay products in the
central rapidity region.

The formalism can be generalised to any colour-singlet hard
sub-process, and the main ingredients to construct the amplitude is the
matrix element for this sub-process and the so-called off-diagonal
unintegrated parton densities. The latter can be interpreted as the
amplitude related to the probability of finding gluons in a proton
with equal but opposite transverse momentum, ${\bf q}_\perp$, and carrying
energy fractions $x$ and $x'$ each, one of which is being probed by a
hard scale $\mu^2$. These densities also include a Sudakov form factor
describing the probability that there is no additional initial-state
radiation from the incoming gluon between the scales $q_\perp$ and
$\mu$, which could destroy the rapidity gaps. Additional emissions
below $q_\perp$ are then suppressed since they cannot resolve the
individual colours of the two gluons. A third ingredient is the
so-called \textit{soft survival probability} which gives the probability that
there are no additional soft or semi-hard interactions between the
colliding protons which could destroy the rapidity gap.

Implementing the Durham formalism for CEP in an event generator is
fairly straightforward since the final states are quite simple and
clean. The cross section for any sub-process can be decomposed in a
central exclusive luminosity function which is folded with a
colour-singlet matrix element in a specific spin state.
Several implementations have been made
\cite{Monk:2005ji,Boonekamp:2011ky,Harland-Lang:2015cta}, and in this
paper will present yet another.

Our implementation is provided as an add-on\footnote{The code uses the
  \pytppp \texttt{UserHooks} machinery and is available on request from
  the authors.} to \pytppp~\cite{Sjostrand:2014zea} and is inspired
by the observation that the Sudakov form factors in the off-diagonal
unintegrated parton densities used within the Durham formalism can be
interpreted in terms of no-emission probabilities in the parton shower
language of \pytppp.

In this way, we can reformulate the cross section for producing a CEP
event in terms of a probability that a standard inclusive sub-process
generated by \pytppp at some scale during the parton shower evolution
is converted to a colour-singlet, and thereafter be considered a CEP
event disallowing further initial-state shower splitting. The main
advantage of this approach is that we actually are allowed to include
initial-state shower splittings, and thus can approximately model
higher order corrections to the original sub-process.
In addition, we have the option of using the multiple interactions
machinery of \pytppp to directly model soft survival probability as
suggested in \cite{Cox:1999dw}.

Consider, \eg, the central exclusive production of di-jets. We would
start by generating the basic $2\to2$ hard partonic scattering from
the inclusive matrix element. We would then generate an initial-state
parton emission from each of the incoming partons. This implicitly
includes the probability that no emission has been made at a higher
scale than the two generated splittings. If the colour and spin state
of the original $2\to2$ is consistent with a CEP, we basically take
the ratio of the corresponding exclusive cross section and the one
calculated with the inclusive matrix element, using the generated scales
as factorisation scale. This gives us the probability to discard the
generated splittings and continue the event as a CEP, or to keep the
hardest emission and continue as a normal inclusive event.

If we continue the event as inclusive, we keep the hardest 
splitting and again generate one initial-state splitting from each
side. Now that we have a three-parton final state which must be
checked if it can be a candidate for CEP, but otherwise the procedure
is repeated. It should be noted that \pytppp in general does not
assign spin states to particles, so the procedure here also involves a
spin decomposition of the parton splitting probabilities and the
matrix elements to correctly get the probability for this to be a
CEP. This will become cumbersome when we go up in parton multiplicity,
but is still fairly straightforward.

The fact that we can stop the parton shower at any stage and check if
we can convert the generated state exclusive, does not only mean that
we can approximate higher order contributions from initial-state
radiation. If we continue the parton shower evolution to low scales
we are also able to investigate the transition region between the
Durham formalism and the resolved Pomeron formalism
\cite{Ingelman:1984ns} which may produce similar final states.

The outline of this paper is as follows. First, we recapitulate the main
features of the Durham formalism in \sectref{sec:durham}. Then we
describe the different parts of our implementation in the \pytppp
program, starting in \sectref{sec:luminosity} with the
reinterpretation of the exclusive luminosity function in terms of the
parton shower no-emission probabilities, and followed by a description
(\sectref{sec:matrix-elements}) of the spin and colour decomposition
of a given partonic state generated by a parton shower from an
inclusive hard matrix element. In \sectref{sec:results} we then
present some proof-of-concept results for some sample processes
before we conclude with a summary and outlook in
\sectref{sec:outlook}.

\section{The Durham formalism}
\label{sec:durham}
Within the Durham model, the amplitude, $\mathcal{A}$, of the central exclusive
process in a $pp$ collision,
\begin{equation}
p p \to p + X + p,
\end{equation}
can be written as
\begin{eqnarray} 
\frac{i\mathcal{A}}{s}&=& \int \frac{\pi^2\, \dd^2 {\bf q}\, \overline{\mathcal{M}}}{{\bf q}^2 ({\bf q}-{\bf p}_{1})^2(-{\bf q}-{\bf p}_{2})^2}\nonumber\\
&&\times f_g(x_1,x_1', Q_1^2,\mu^2;t_1)f_g(x_2,x_2',Q_2^2,\mu^2;t_2) \, ,
\label{eq:DurhamAmplitude}
\end{eqnarray}
where the integration runs over the two-dimensional transverse momentum of the screening gluon ${\bf q}$ (Fig.~\ref{fig:exdiagram}).
The transverse momenta of the outgoing protons are denoted as ${\bf p}_{1}$ and ${\bf p}_{2}$.
The scales $Q_1^2$ and $Q_2^2$ are within Durham model \cite{Harland-Lang:2013ncy} defined as
\begin{eqnarray}
Q_1^2 &=& \min \left[\, {\bf q}^2,\,({\bf q} - {\bf p}_{1})^2\, \right], \\
Q_2^2 &=& \min \left[\,{\bf q}^2,\,(-{\bf q} - {\bf p}_{2})^2\, \right],
\end{eqnarray}
\ie\ as the smaller of the virtualities of the screening gluon and the
fusing gluon momenta related to the particular proton.

The proton form factors, $F_N$, are absorbed into off-diagonal unintegrated PDFs $f_g$.
These PDFs are assumed to factorise as:
\begin{equation}
f_g(x, x', Q^2,\mu^2;t) = f_g(x,x', Q^2,\mu^2) F_N (t),
\end{equation}
where $t_{1,2} \approx - {\bf p}_{1,2}^2$. The proton form
factors in the simplest approach are $F_N(t) = \eu^{bt/2}$, with $b
\approx 4\,\GeV^{-2}$.  These off-diagonal unintegrated
densities depend on the momentum fraction of the fusing (screening)
gluon $x$ ($x'$) and on two scales: the scale of the hard sub-process
$\mu^2$; and the scale corresponding to the screening gluon transverse
momentum $Q^2$.

The kinematic regime relevant for CEP is $Q/\sqrt{s}\sim x' \ll x\sim
M_X/\sqrt{s}$, which allows to integrate out the $x'$-dependency of
$f_g$ and express them using generalised gluon PDFs $H_g$ and the Sudakov
factor \Tsud:
\begin{eqnarray}
f_g(x,x', Q^2, \mu^2) &=& \nonumber \\
&&\hspace{-4em}\frac{\partial }{\partial \ln Q^2 } \left[ H_g \left(\frac{x}{2}, \frac{x}{2}; Q^2 \right) \sqrt{\Tsud(Q^2, \mu^2)} \right]
\label{UnOffGluon}
\end{eqnarray}
The Sudakov factor \Tsud\ describes the probability of no emission
from the fusing gluons between scales $Q^2$ and $\mu^2$.  It resumes
singularities from virtual diagrams with soft or collinear emissions
up to (modified) next-to-leading logarithmic accuracy and ensures that
the integral (\ref{eq:DurhamAmplitude}) is finite, as the Sudakov
factors exponentially suppress low ${\bf q}$ contribution.
\begin{eqnarray}
\Tsud(Q^2, \mu^2) &=& \exp \bigg( - \int_{Q^2}^{\mu^2} \frac{\dd k^2}{k^2} \frac{\alpha_s (k^2)}{2\pi} \nonumber\\
&&\hspace{-2em}\times \int_{0}^{1-\epsilon(k/M_X) }\! \dd z\, \left[ z P_{gg}(z) + n_f P_{qg} (z) \right] \bigg)
\label{eq:SudakovFactor}
\end{eqnarray}
In this expression both the splitting functions $P_{gg}$, $P_{qg}$ and
the running of $\alpha_s$ are in the leading order form.
The upper bound of
the $z$ integration\footnote{Usually simply $\epsilon(k/M_X) =
  k/M_X$.} depends on the mass of the exclusive system which
makes the Sudakov factor \Tsud\ also $M_X$-dependent as indicated by
the subscript, $M$.

The generalised PDF $H_g$ \cite{Harland-Lang:2013xba} can be
approximately calculated from the ordinary parton distribution
function of gluon $g(x,Q^2)$ using relation:
\begin{equation}
H_g \left(\frac{x}{2}, \frac{x}{2};Q^2 \right) = \frac{4}{\pi} \int_{\frac{x}{4}}^{1}\dd y\; x \sqrt{y(1-y)} \;  g \left( \frac{x}{4y}, Q^2 \right),
\label{eq:HgFormula}
\end{equation}
In a much used approximation the generalised PDF, $H_g$, is simply
proportional to the conventional one
\begin{equation}
H_g \left(\frac{x}{2}, \frac{x}{2};Q^2 \right) =  R_g\, x\, g(x,Q^2),
\end{equation}
where the constant $R_g$ is about $1.3$ for LHC energies \cite{Harland-Lang:2013xba}.

The sub-process amplitude $\mathcal{M}$ depends on transverse momenta
of the fusing gluons ${\bf q}_1$ and ${\bf q}_2$
\begin{equation}
{\bf q}_1 = {\bf q} - {\bf p}_{1} \qquad  {\bf q}_2 = -{\bf q} - {\bf p}_{2} 
\end{equation}
and on the $gg \to X$ vertex $V_{ij}^{ab}$ which is averaged over
identical colour indexes $a=b$.

\begin{equation}
\mathcal{M} = \frac{2}{M_X^2} \frac{1}{N_C^2-1} \delta^{ab} q_{1}^i q_2^j V^{ab}_{ij}
\end{equation}
The sub-process amplitude is typically known in the helicity basis. In this basis the vertex term $q_1^i q_2^j \, V^{aa}_{ij}$ takes form:
\begin{eqnarray}
  q_1^i q_2^j \, V^{aa}_{ij} &=&\nonumber\\
&&\hspace{-3em}=\left\{
    \begin{array}{ll}
      -\frac{1}{2} (q^x_1 q^x_2 + q^y_1 q^y_2  ) 
      \times(\Amp_{++} + \Amp_{--} ) \\
      -\frac{i}{2} (q^x_1 q^y_2 - q^y_1 q^x_2 ) 
      \times(\Amp_{++} - \Amp_{--} ) \\
      +\frac{1}{2}\left[(q^x_1 q^x_2 - q^y_1 q^y_2) +
        i(q^x_1 q^y_2 + q^y_1 q^x_2)\right] \times \Amp_{-+} \\
      +\frac{1}{2}\left[(q^x_1 q^x_2 - q^y_1 q^y_2) -
        i (q^x_1 q^y_2 + q^y_1 q^x_2 )\right]  \times\Amp_{+-} 
    \end{array} \right. \nonumber\\
&&\hspace{-3em}\equiv\left\{
    \begin{array}{ll}
      \spinkin_{0^+}({\bf q}_1,{\bf q}_2) & \times(\Amp_{++} + \Amp_{--} )  \\
      \spinkin_{0^-}({\bf q}_1, {\bf q}_2)& \times(\Amp_{++} - \Amp_{--} )\\
      \spinkin_{+2^+}({\bf q}_1, {\bf q}_2) &\times \Amp_{-+}\\
      \spinkin_{-2^+}({\bf q}_1,{\bf q}_2) & \times\Amp_{+-}
    \end{array} \right. ,
  \label{eq:spinkin}
\end{eqnarray}
where we have introduced the kinematic spin factors
$\spinkin_{J_z}({\bf q}_1,{\bf q}_2)$ for future reference with $q_1^i = q^i
- p_{1}^i$ and $q_2^i = -q^i - p_{2}^i$ ($i=x,y$),
and the amplitudes, \eg\ $\Amp_{++}$, depend on $M_X$, the momenta of
outgoing particles, the helicities of outgoing particles as well as
the helicities of incoming gluons (here both $+1$).  The amplitudes
also depend on the colours of the particles in the sub-process.
Here, the amplitudes $\Amp$ are the result of averaging over colour
indexes of the incoming partons in such a way that the exclusive
system is a colour singlet.

It is useful to note that in the collinear limit, where ${\bf q}_1 =
-{\bf q}_2 = {\bf q}$ the kinematic factors are simply:
\begin{eqnarray}
  \label{eq:collimit}
   \spinkin_{0^+}({\bf q},-{\bf q})=\frac{1}{2} {\bf q}^2, \qquad \spinkin_{0^-}({\bf q},-{\bf q})=0,\nonumber\\
  \mbox{and}\qquad
   \spinkin_{\pm 2^+}({\bf q},-{\bf q})=-\frac{1}{2} {\bf q}^2\, \eu^{\pm 2i\phi},
\end{eqnarray}
where the azimuthal angle of ${\bf q}$ was labelled as $\phi$. The
state $J_z=0^-$ is trivially zero as a consequence of cross product of
two collinear vectors.  States $|J_z|=2$ become zero after integration
over $\phi$ because the remaining part of (\ref{eq:DurhamAmplitude})
is $\phi$ independent in the collinear limit.  Beyond such collinear
limit also non spin-singlet states contribute to the cross section,
but are suppressed as $\langle {\bf q}^2 \rangle^2 / \langle {\bf p}_{1,2}^2
\rangle^2 \sim 0.01$ with respect to the spin singlet term
\cite{Harland-Lang:2015cta} (if all helicity amplitudes, $\Amp$, are of
the same size).

Note that the formula (\ref{eq:DurhamAmplitude}) does not include a
soft survival probability which cannot be calculated in a perturbative
way.  It reduces the CEP cross section at LHC by
typically two orders of magnitude and can, \eg, be determined using the
eikonal model \cite{Harland-Lang:2013xba,Khoze:2014aca,Khoze:2002nf}.

\section{Reinterpretation of the exclusive cross section}
\label{sec:luminosity}

\subsection{Prerequisites}
In the following discussion it will be beneficial to reformulate the
CEP amplitude of the Durham model into more straightforward form and
rather work directly with the exclusive cross section $\sigma^{exc}$
\begin{eqnarray}
\sigma^{exc} &=& \int \dd y\, \dd\ln M^2  \dd^2{{\bf p}_1}  \dd^2{\bf p}_{2}\, \eu^{-b {\bf p}_1^2}
\eu^{-b {\bf p}_2^2}  \frac{1}{64\pi^2} \frac{1}{2M^2} \nonumber\\
&\times&\quad \bigg| \int \frac{\dd^2 {\bf q}\,\, q_1^i q_2^j \, V_{ij}^{aa} (M,w, \mu_R^2)}{{\bf q}^2 ({\bf q}-{\bf p}_1)^2 (-{\bf q}-{\bf p}_2)^2 }\label{MasterExclusiveAmplitude} \\ 
&\times& f_g^M (x_1, x'_1, Q_1^2, \mu_F^2) f_g^M (x_2, x'_2, Q_2^2, \mu_F^2)\;  \bigg|^2 \dd\Sigma_w. \nonumber 
\end{eqnarray}
The variables $y$ and $M$ denote the rapidity and mass of the exclusive system and are related to the momenta fractions $x_1$ and $x_2$ by formulas
\begin{equation}
x_{1}=\frac{M} {\sqrt{s}}\; e^{y} \qquad  x_{2}= \frac{M} {\sqrt{s}}\; e^{-y}.
\end{equation}
The transverse momenta ${\bf p}_1$ and ${\bf p}_2$ of the scattered protons are
assumed to be distributed according the simple one-channel model with
the slope of the exponential equal to $b \sim 4\,\GeV^{-2}$.  The
integration inside the absolute value is performed over transverse
momentum ${\bf q}$ of the ``screening'' gluon.  The other variables are
recognised from the previous section.  For completeness the dependency
of the off-diagonal generalised PDFs on mass $M$ (via a Sudakov form
factor) as well as the arguments of the $V_{ij}^{aa}$ are written
explicitly.  The kinematics of all outgoing particles of the exclusive
system $X$ is denoted by $w$ and $\dd\Sigma_w$ is the corresponding
phase space element.  The whole expression is integrated over phase
space of these outgoing momenta $w$ which satisfy imposed kinematic
cuts.

Inspired by a similar form of the derivative of the exponential
function, we factorise the Sudakov factor in expression
(\ref{UnOffGluon}) in front of the bracket which leads to
\begin{equation}
 f_g^M(x,x', Q^2, \mu_F^2) = \sqrt{\Tsud(Q^2,\mu_F^2)} \;
    \phi_M (x, Q^2),
\end{equation}
where the newly introduced modified PDF $\phi_M$ is defined as:
\begin{eqnarray}
\phi_M (x, Q^2) &=&  \bigg[ \frac{\dd H_g(\frac{x}{2},\frac{x}{2}; Q^2)}{\dd\ln Q^2} \\
     &&+\frac{\alpha_s(Q^2)}{4\pi} H_g \left(\frac{x}{2},\frac{x}{2}; Q^2\right) \nonumber \\
&&\times \int_{0}^{1-\epsilon\left(\frac{Q}{M}\right)}
  \dd z\left[zP_{gg}(z) + n_f(Q) P_{qg}(z)\right]      \bigg] \nonumber
\end{eqnarray}

In the text below, for simplicity, only the dominant spin singlet part
will be considered and the following abbreviation is introduced:
\begin{equation}
\mathcal{D}^2_{1,2}\, {\bf q} = \frac{\dd^2 {\bf q}} {{\bf q}^2 ({\bf q}-{\bf p}_1)^2 (-{\bf q}-{\bf p}_2)^2 } ({\bf q}-{\bf p}_1) \cdot (-{\bf q} - {\bf p}_2),
\end{equation}
where the indexes $1,2$ indicate the dependency of the differential on
transverse momenta ${\bf p}_1$ and ${\bf p}_2$ and the dot represents the scalar
product of two-dimensional vectors. In the collinear
limit\footnote{\ie\ $|{\bf p}_1| \ll |{\bf q}|$ and $|{\bf p}_2| \ll |{\bf q}|$.}
this differential simplifies to $- \dd^2 {\bf q} / {\bf q}^4$.

Using these notations and assumptions the formula
(\ref{MasterExclusiveAmplitude}) takes form:
\begin{eqnarray}
  \sigma^{exc} &=& \int \dd y\, \dd\ln M^2  \dd^2{{\bf p}_1}  \dd^2{\bf p}_{2}\,
  \eu^{-b {\bf p}_1^2}  \eu^{-b {\bf p}_2^2} \nonumber\\
  &&\hspace{-3.3em}\times\frac{1}{256\pi^2} \frac{1}{2M^2} 
  \mathcal{D}^2_{1,2}\, {\bf q} \;  \mathcal{D}^2_{1,2}\, {\bf q}' \nonumber    \\
  &&\hspace{-3.3em}\times\phi_M(x_1,Q_1^2)\phi_M(x_2,Q_2^2)\phi_M(x_1,Q_1'^2)\phi_M(x_2,Q_2'^2)
  \nonumber \\
  &&\hspace{-3.3em}\times\Tsud^{1/2}(Q_1^2,\mu_F^2) \Tsud^{1/2}(Q_2^2,\mu_F^2)
  \Tsud^{1/2}(Q_1'^2,\mu_F^2)\Tsud^{1/2}(Q_2'^2,\mu_F^2)\nonumber \\
  &&\hspace{-3.3em}\times|\Amp_{++}+\Amp_{--}|^2  \dd \Sigma_w,
  \label{eq:ExclCrossSection}
\end{eqnarray}
where ${\bf q}'$ is the transverse momentum of the screening gluon in the complex conjugate amplitude.

In the next step, the exclusive cross section
(\ref{eq:ExclCrossSection}) will be expressed as a product of the
exclusive luminosity and the exclusive cross section.  We define the
spin-singlet colour-singlet cross section of hard sub-process as:
\begin{equation}
  \sigma^s (w,\mu_r^2) = \frac{1}{4} \frac{1}{64} 
  \frac{1}{2M^2} \sum_{\lambda, j}  \left| \sum_{a=1}^{8}
    \left( \Amp_{++\to \lambda}^{aa \to j}+
      \Amp_{-- \to \lambda}^{aa \to j} \right) \right|^2,
\label{eq:spinSingletColSingletXsec}
\end{equation}
where the factor $1/4$ follows from the probability to have some
particular helicity configuration of the incoming gluons and the
coefficient $1/64$ has an analogous meaning for the colours.
The term $1/2M^2$ represents the ``flux factor''.
The symmetrisation factor, $N_s$, important if identical particles
occur in the final state, is assumed to be incorporated in
the amplitudes $\Amp$, \ie\ amplitudes are scaled by $1/\sqrt{N_s}$.
Letters $\lambda$ and $j$ denote all possible helicity and colour
configurations of the exclusive system.

The exclusive luminosity which corresponds to the exclusive cross
section of the hard sub-process (\ref{eq:spinSingletColSingletXsec})
is:
\begin{eqnarray}
  L_{exc}(M,y,\mu_F^2) &=& \\
  &&\hspace{-5em}\times\int \frac{1}{\pi^2}  \dd^2{{\bf p}_1}  \dd^2{\bf p}_{2}\,
  \eu^{-b {\bf p}_1^2}\,  \eu^{-b {\bf p}_2^2}\,
  \mathcal{D}^2_{1,2}\, {\bf q}  \; \mathcal{D}^2_{1,2}\, {\bf q}'\nonumber\\
  &&\hspace{-5em}\times\phi_M(x_1,Q_1^2)\phi_M(x_2,Q_2^2)\phi_M (x_1,Q_1'^2)\phi_M (x_2,Q_2'^2)
  \nonumber\\
  &&\hspace{-5em}\times\Tsud^{1/2}(Q_1^2,\mu_F^2)\Tsud^{1/2}(Q_2^2,\mu_F^2) \nonumber\\
  &&\hspace{-5em}\times\Tsud^{1/2}(Q_1'^2,\mu_F^2)\Tsud^{1/2}(Q_2'^2,\mu_F^2). \nonumber
\end{eqnarray}
In the collinear limit, where ${\bf p}_1$ and ${\bf p}_2$ are neglected with
respect to ${\bf q}$ and ${\bf q}'$, the luminosity can be integrated over ${\bf p}_1$
and ${\bf p}_2$ and over the azimuthal angles of ${\bf q}$ and ${\bf q}'$ which leads
to:
\begin{eqnarray}
  L_{exc}(M,y,\mu_F^2) &=& \frac{\pi^2}{b^2} \int\!
  \frac{\dd q^2}{q^4} \frac{\dd q'^2}{q'^4}  \\
  &&\hspace{-4em}\times \phi_M (x_1,q^2)\phi_M(x_2,q^2)\phi_M (x_1,q'^2)\phi_M (x_2,q'^2) \nonumber \\
  &&\hspace{-4em}\times \Tsud(q^2,\mu_F^2)  \Tsud(q'^2,\mu_F^2) \nonumber
\label{KMRintegrand2D}
\end{eqnarray}
\begin{eqnarray}
L_{exc} (M,y,\mu_F^2) &=&\\
 &&\hspace{-5em}\frac{\pi^2}{b^2} \bigg| \int    \frac{\dd\ln q^2}{q^2} 
\phi_M (x_1,q^2)  \phi_M (x_2,q^2) \; \Tsud(q^2,\mu_F^2) \bigg|^2 \nonumber
\label{KMRintegrand}
\end{eqnarray}
The integrand in (\ref{KMRintegrand2D}) is fairly flat if considered as a function of $1/q^2$
and $1/q'^2$ which makes these variables suitable for Monte Carlo
integration\footnote{The integral over $k_\perp$ in the definition
  of the Sudakov factor \Tsud\ (\ref{eq:SudakovFactor}) can be
  evaluated numerically by means of Gauss-Kronrod quadrature formula \cite{Kronrod:1964}.
  The relative error of \Tsud\ is then
  typically $10^{-16}$ if the function values in 15 points are used
  for the numerical integration.}, 
  especially because $\dd (1/q^2) = -\dd q^2/q^4$.

\begin{figure*}
  \begin{minipage}{0.49\textwidth}
    \begin{center}
      \includegraphics[width=\linewidth]{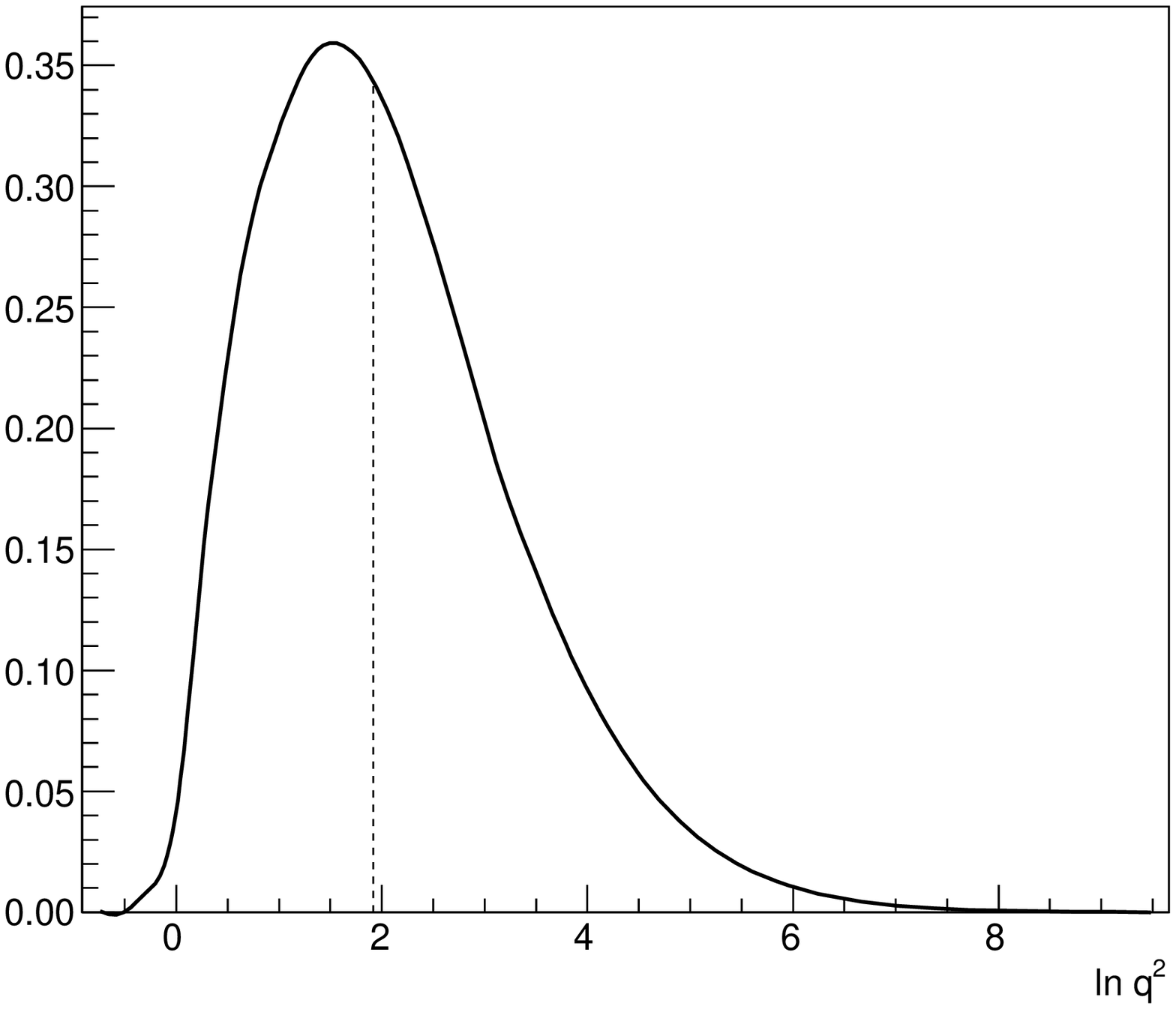}
    \end{center}
  \end{minipage}
  \begin{minipage}{0.49\textwidth}
    \begin{center}
      \includegraphics[width=\linewidth]{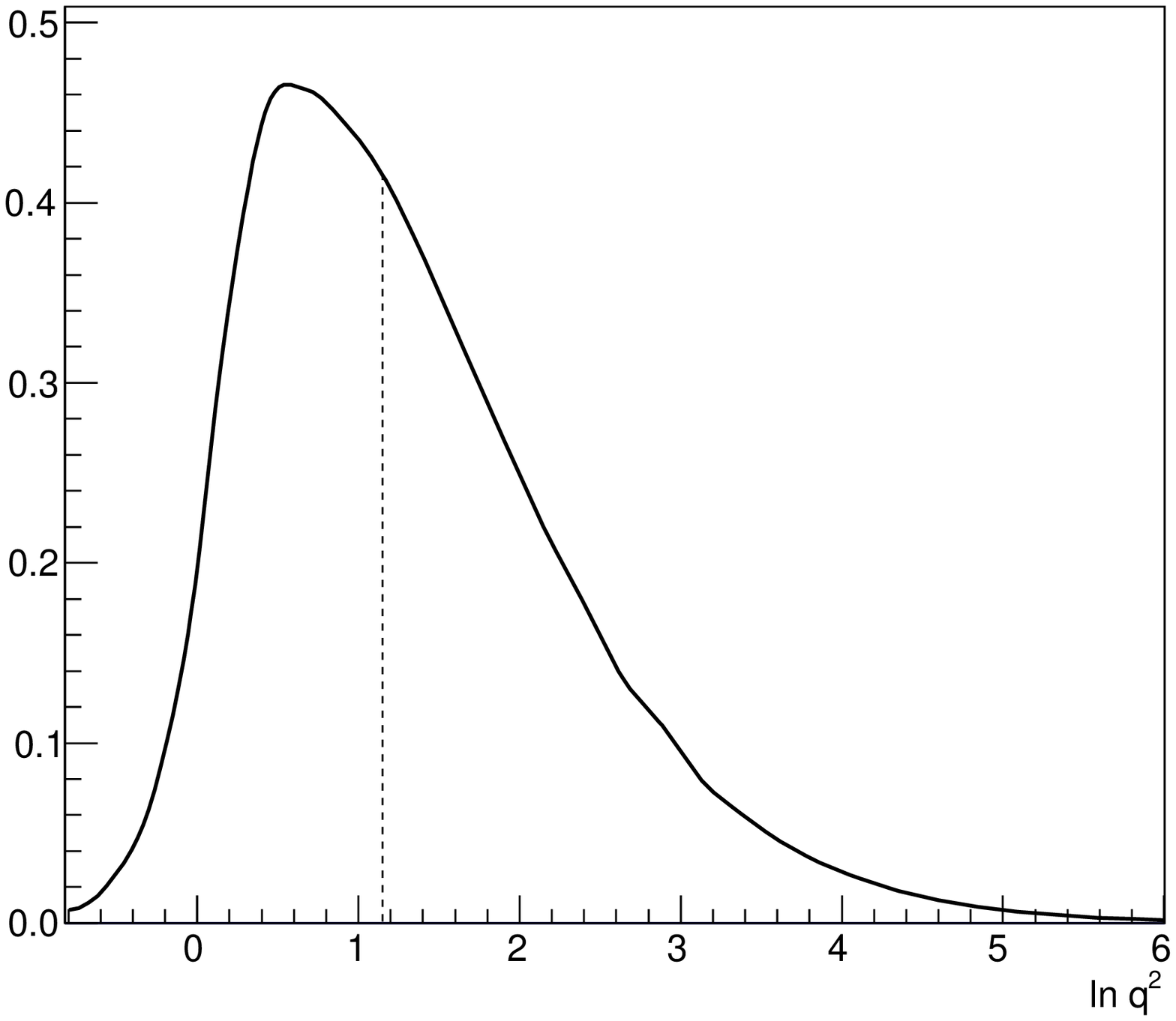}
    \end{center}
  \end{minipage}
  \caption{  Integrand of formula
    (\ref{KMRintegrand}) for Higgs production at LHC,
    $\sqrt{s}=13\,\TeV$ and $\mu_F=125\,\GeV$ (left), and for di-jet
    production at Tevatron, $\sqrt{s}=1.96\,\TeV$ and $\mu_F=25\,\GeV$
    (right). The dashed lines depict the medians of the
    distributions, which are $2.6\,\GeV$ for LHC and $1.8\,\GeV$ for Tevatron.
    We calculated the generalised gluon PDF $H_g$ from MMHT2014 LO PDF
		using formula~(\ref{eq:HgFormula}).}
  \label{fig:Integrant}
\end{figure*}

The integrand in (\ref{KMRintegrand}) is shown in
Fig.~\ref{fig:Integrant} for $\ln q^2$ as an integration variable.
For calculating of the exclusive luminosity the upper limit of
integration is set to the factorisation scale $\mu_F$, nevertheless
the integrand is typically negligible in the high $q$ region.  Whereas
it dominates for $q$ of $2-3\, \GeV$ in case of LHC Higgs production
and for even smaller values ($1-2\, \GeV$) in case of di-jet production at Tevatron.
Values below the starting scale $q_0 = 1\, \GeV$ of MMHT2014 LO PDF
\cite{Harland-Lang:2014zoa} can be extracted using a backward DGLAP
evolution \cite{Botje:2010ay}.  In reality, it was argued in
\cite{Khoze:2004yb} that for $q \lesssim 0.85\, \GeV$ the gluon
propagator would be modified by non-perturbative dynamics which
effectively suppress such low $q$ contributions.  Rather than a sharp cut-off, the
damped gluon PDF \cite{Harland-Lang:2015cta} is used to calculate
$\phi_M$ below $q_0$ in order to suppress the region of low transverse
momentum:
\begin{eqnarray}
g(x,q) = g(x,q_0) \left(\frac{q^2}{q_0^2}\right)^{2+(a_1-2) (q^2/q_0^2) + a_2 (q^2/q_0^2)^2 } \nonumber\\\text{for}\quad q<q_0
\end{eqnarray}
The coefficients $a_{1,2}$ are chosen in such a way that the function is smooth in $q_0$ up to the second derivative.

Finally, the CEP cross section expressed as a convolution of the exclusive luminosity (\ref{KMRintegrand}) and the exclusive cross section of the hard sub-process (\ref{eq:spinSingletColSingletXsec}) takes from
\begin{equation}
\sigma^{exc} = \int \dd y\, \dd\ln M^2 \; L_{exc}(M,y,\mu_F^2)\,  \sigma^s(M, w, \mu_R^2)\,  \dd\Sigma_w,
\label{exclusive_noSel}
\end{equation}
in analogy with the formula formula for the inclusive cross section
\begin{equation}
\sigma^{inc} = \int \dd y\, \dd\ln M^2 \; L_{inc}(M,y,\mu_F^2)\,  \sigma^i(M, w, \mu_R^2)\,  \dd\Sigma_w,
\label{inclusive_noSel}
\end{equation}
where the inclusive luminosity $L_{inc}$ is
\begin{equation}
L_{inc}(M,y,\mu_F^2) = x_1 g(x_1,\mu_F^2)\;  x_2 g(x_2,\mu_F^2).
\end{equation}

\subsection{Screening gluons in the \protect\pytppp interleaved parton shower}
\label{sec:approach1}

The way parton showers are included in \pytppp is through an
interleaved process where we normally have three competing processes,
which we will denote ISR, FSR and MPI. ISR is an initial-state
splitting where one of the incoming partons to the hard sub-process is
evolved to lower scale and higher energy fraction, by emitting a
parton into the final state; FSR is the final-state splitting of a
parton in the final state; while MPI is the appearance of an
additional parton--parton interaction. All of these occur at
decreasing scale, where the highest scale is given by the kinematics
of the hard sub-process. In each step in the shower we then pick a
process which has a lower scale than the previous one, and the
factorisation property of the no-emission probability means that we
can generate one of each of the possible processes independently and
simply pick the one which yielded the highest scale in each step.

For simplicity we will here only consider the ISR, concentrating on
the initial-state $g\to gg$ splittings, and show how we can
reinterpret the formula for CEP as an extra process in the interleaved
shower, which transforms an inclusive event into an exclusive one.

First, let's consider the inclusive processes with no emission from
the space-like shower between scales $\mu^2$ and $\mu_F^2$.  The scale
$\mu_F^2$ is considered as a starting scale of the backward space-like
parton shower.  The cross section for such processes take form:
\begin{eqnarray}
  \sigma^{inc}(\mu^2) &=& \int \dd y\, \dd\ln M^2 \; L_{inc}(M,y,\mu_F^2)\\
  &&\Tnoem{x_1}(\mu^2, \mu_F^2)  \Tnoem{x_2}(\mu^2, \mu_F^2)
  \sigma^i(M, w, \mu_r^2)\,  \dd\Sigma_w, \nonumber
\end{eqnarray}
where \Tnoem{x} is the no-emission probability, quantifying the
probability that no extra emission from parton are present between two
given scales, if the higher scale is taken as a reference.  This
no-emission term is used in the backward evolution of the space-like
showers in Pythia and, for the case of an incoming gluon, it is
defined as:
\begin{eqnarray}
  \Tnoem{x}( q_1^2, q_2^2) &=& \exp \bigg( - \int_{q_1^2}^{q_2^2}
    \frac{\dd q^2}{q^2} \frac{\alpha_s (q^2)}{2\pi} \\
    &&\times\sum_a \int_{x}^{1-\epsilon \left(\frac{q}{M}\right)}
    \frac{\dd z}{z} \frac{f_a (\frac{x}{z}, q^2)}{g(x,q^2)} P_{a\to g}(z)\bigg),\nonumber
\end{eqnarray}
where the sum runs over gluons and all possible flavours of quarks and
anti-quarks.  It can be shown that these no-emission probabilities are
linked to the standard Sudakov form factors by the relation
\cite{Krauss:2002up,Lavesson:2005xu}:
\begin{equation}
  \Tnoem{x}( q_1^2, q_2^2) = 
  \frac{ g(x,q_1^2)}{ g(x,q_2^2) } \Tsud(q_1^2, q_2^2)
\label{Sudakov_forw_back}
\end{equation}

The cross section $\sigma^{inc}$, differential in the variable $\ln
\mu^2$, corresponding to the scale of the first parton shower emission
is:
\begin{eqnarray}
  \frac{\dd \sigma^{inc}(\mu^2)}{\dd \ln \mu^2} = \int \dd y\, \dd\ln M^2 \;
  L_{inc}(M,y,\mu_F^2)\times\nonumber\\
   \frac{\dd}{\dd \ln \mu^2} 
  \Big(\Tnoem{x_1}(\mu^2, \mu_F^2) \Tnoem{x_2}(\mu^2, \mu_F^2) \Big) \nonumber\\
  \sigma^i(M, w, \mu_R^2)\,  \dd\Sigma_w
\label{inc_sigma_1em_bw}
\end{eqnarray}

Employing the relation (\ref{Sudakov_forw_back}) the derivative of the
no-emission probabilities can be expressed using the Sudakov factors
only:
\begin{eqnarray}
  \frac{\dd}{\dd\ln \mu^2} \Big( \Tnoem{x_1} (\mu^2, \mu_F^2) 
  \Tnoem{x_2} (\mu^2, \mu_F^2)\Big) = \Tsud(\mu^2,\mu_F^2)^2\\
  \times\frac{\tilde{g}(x_1,\mu^2)x_2 g(x_2,\mu^2)+
    x_1 g(x_1,\mu^2)\tilde{g}(x_2,\mu^2)}
  {x_1 g(x_1,\mu_F^2) x_2 g(x_2,\mu_F^2)},\nonumber
\end{eqnarray}
where the newly defined distribution function $\tilde{g}$
is:\footnote{The function $\tilde{g}$, resembling $\phi_M$, depends also on mass $M$.}
\begin{eqnarray}
  \tilde{g} (x,\mu^2) &=&
  x\frac{\partial g(x,\mu^2)}{\partial \ln q^2} + x g(x,\mu^2) \\
  &&\hspace{-2em}\times\,\frac{\alpha_s(\mu^2)}{2\pi} \int_0^{1-\epsilon(\mu/M)} \dd z
  \left[zP_{gg}(z) + n_f(\mu) P_{qg}(z)\right]  \nonumber
\end{eqnarray}
It allows to re-express the differential cross section
(\ref{inc_sigma_1em_bw}) using the standard Sudakov form factors:
\begin{eqnarray}
  \frac{\dd \sigma^{inc}(\mu^2)}{\dd \ln \mu^2} &=&
  \int \dd y\, \dd\ln M^2 \\
  &&\hspace{-3em}  \Big[ \tilde{g}(x_1,\mu^2) x_2 g(x_2,\mu^2) +
    x_1 g(x_1,\mu^2) \tilde{g}(x_2,\mu^2)  \Big]\nonumber\\
  &&\hspace{-3em}\qquad\quad\times\;\Tsud(\mu^2,\mu_F^2)^2\, \sigma^i(M, w, \mu_R^2)\,  \dd\Sigma_w \nonumber
\label{inc_sigma_1em_fw}
\end{eqnarray}
The inclusive cross section is then simply:
\begin{equation}
\sigma^{inc} = \int^{\mu_F^2} \dd \ln \mu^2 \,  \frac{\dd \sigma^{inc}(\mu^2)}{\dd \ln \mu^2},
\end{equation}
where the lower integration limit is assumed to be so small that the Sudakov factor
between this limit and $\mu_F^2$ is close to zero.

We can now apply a similar procedure to the exclusive cross section
where the variable $\mu^2$ is now interpreted as the minimal
transverse momentum of the screening gluons exchanged during the
interaction.  Let's define:
\begin{eqnarray}
  \frac{\dd L_{exc}(M,y,\mu_F^2,\mu^2)}{\dd \ln \mu^2} &=&
  \frac{\dd}{\dd \ln \mu^2} \frac{\pi^2}{b^2}\\
  &&\hspace{-9.5em}\times\;\bigg| \int_{\mu^2}^{\mu_F^2}   \dd \ln q^2 \; \frac{1}{q^2} 
    \phi_M (x_1,q^2)  \phi_M (x_2,q^2) \; \Tsud(q^2,\mu_F^2) \bigg|^2\nonumber \\
  &&\hspace{-8em}= - \frac{\pi^2}{b^2} \frac{2}{\mu^2} \phi_M (x_1,\mu^2)
  \phi_M (x_2,\mu^2) \Tsud(\mu^2,\mu_F^2)^2\times\nonumber\\
  &&\hspace{-7em}\times\int_{\mu^2}^{\mu_F^2} \frac{\dd \ln q^2}{q^2} \phi_M (x_1,q^2)  \phi_M (x_2,q^2) \frac{1}{\Tsud(\mu^2,q^2)}\nonumber
\label{lumi_exc_diff}
\end{eqnarray}
Then the derivative of the exclusive cross section according to this
variable is simply:
\begin{eqnarray}
  \frac{\dd \sigma^{exc}(\mu^2)}{\dd \ln \mu^2} &=& \int \dd y\, \dd\ln M^2 \nonumber\\
  &&\hspace{-1.3em}\frac{\dd L_{exc}(M,y,\mu_F^2,\mu^2)}{\dd \ln \mu^2} \, 
  \sigma^s(M, w, \mu_R^2)\,  \dd\Sigma_w.
\label{exc_sigma_derivative}
\end{eqnarray}
The ratio of the integrands in \eqsref{exc_sigma_derivative}
and \eq{inc_sigma_1em_fw} defines the exclusive probability
\begin{eqnarray}
  p_{exc} &=& \frac{-\frac{1}{\Tsud(\mu^2,\mu_F^2)^2}
  \frac{\dd L_{exc}(M,y,\mu_F^2,\mu^2)}{\dd\ln\mu^2}}  
  {\left( \tilde{g}(x_1,\mu^2) x_2 g(x_2,\mu^2) + 
      x_1 g(x_1,\mu^2) \tilde{g}(x_2,\mu^2)  \right) }\nonumber\\
  &&\times\frac{\sigma^s(M, w, \mu_R^2)}{\sigma^i(M, w, \mu_R^2)}.
  \label{eq:pexc}
\end{eqnarray}
This means that we now have for each step in the interleaved shower a
probability for a given partonic state to become exclusive by the
exchange of a screening gluon. The physical picture is the same as in
the Durham model, in that a screening gluon with low transverse
momentum cannot affect the colour structure of an emission at a higher
scale. It also has the nice property that we become somewhat
insensitive to the low transverse momentum behaviour of the parton
densities in the integral of the exclusive luminosity.

There is, however, a problem with this approach, in that the $p_{exc}$
is very peaked at small $\mu$, making the generation of
the exclusive events very inefficient. Although a weighting procedure could
be applied, it would be difficult to incorporate into the current
framework of \pytppp. For the purpose of this paper, we have therefore
chosen to implement a simpler procedure.

\subsection{A simpler approach}
\label{sec:approach2}

Instead of adding the exchange of a screening gluon as an extra
process in the interleaved shower, we simply calculate before each
shower step, if given state should be made exclusive, using the
probability

\begin{equation}
  p'_{exc} = \frac{L_{exc}(M,y,\mu^2)}{L_{inc}(M,y,\mu^2)}\;
  \frac{\sigma'^s(M, w, \mu_R^2)}{\sigma'^i(M, w, \mu_R^2)},
\label{eq:pexmod}
\end{equation}
where $\mu^2$ is the scale of the latest emission. We note that the
integration in $L_{exc}$ now goes down to very low transverse
momenta, but it turns out that the results are the same as in the more
complicated approach above.

The modified cross section $\sigma'^i$ is defined as:
\begin{equation}
\sigma'^i = P_n(z_{n}) P_{n-1}(z_{n-1})\ldots P_2(z_2) P_1(z_1) \sigma^i,
\label{eq:sigmaMod}
\end{equation}
where the $P_i$ splitting functions in principle could be either
initial- or final-state splittings.  For modified singlet sub-process
cross section the definition is the same, only the formula is
corrected for the fact that the incoming gluons are in colour and spin
singlet state which, for example, makes $H+\mathrm{jet}$ exclusive
cross section equal to zero.  Note that for calculation of the
modified singlet cross section, not only classical matrix element
squared but also amplitudes for all possible helicity configuration
must be known.  The procedure for calculating $\sigma'_s$ will be
provided in the next section.

The full procedure would be to start the generation of an inclusive
process in \pytppp and calculate the probability in
\eqref{eq:pexmod} of that process being exclusive. Then, after each
ISR or FSR step in the interleaved shower, we would again check if the
current state can be made exclusive by \eqref{eq:pexmod}. If the
event is to be considered to be exclusive we would rearrange the
colour flow accordingly and insert the quasi-elastically scattered
protons, but let the shower continue without the ISR process.

Note that the soft survival probability, which we have so far left out of
the exclusive luminosity function, corresponds exactly to the
probability of having no additional multi-parton interactions, so any
MPI in the interleaved shower (before or after the event has been made
exclusive) will mean that the event will stay inclusive.

To make the generation of exclusive events more efficient we have here
decided to simplify the procedure even more. A final-state emission
does not modify the parton densities used in the luminosity functions,
and although they may affect the exclusive cross section, we have here
decided to leave them out and generate them separately.
Also the calculation of the soft survival
probability by vetoing any exclusive event in case the shower gives a
MPI is extremely inefficient, and we instead calculate that separately
and simply multiply the inclusive cross section with that factor\footnote{
In reality we calculate the exclusive cross sections with 
MPI switched on in several bins of the calculated
variables to control possible kinematic dependence of the 
soft survival probability.}.
It should be noted that using the MPI model in \pytppp presented in
\cite{Cox:1999dw} has not been carefully investigated. It has the
interesting feature that the probability for additional scattering
depends on the hardness of the primary scattering, since harder
processes have larger overlap (smaller impact parameter). It also has a
natural dependence on the collision energy. The downside is that it is
sensitive to the soft behaviour of MPI model and may vary strongly
between different tunings of the parameters. In this paper we will
simply use the default tune in \pytppp and postpone a proper
investigation of the procedure to a future publication.

In the end, the procedure will look as follows:

\begin{enumerate}
\item Generate the hard sub-process of interest in \pytppp, use the
  standard inclusive cross section.
\item Use only the initial-state shower in \pytppp.
\item Before generating the next initial-state emission, make the
  event exclusive with the probability in \eqref{eq:pexmod}.
\item If the event is made exclusive, switch off the initial-state
  cascade, rearrange colours, remove the proton remnants, insert the
  scattered protons and continue with final-state radiation from the
  exclusive state.
\item If the event stays inclusive, generate the next initial-state
  emission (continue with step 3).
\end{enumerate}

As an alternative, to enable detailed studies of the exclusive
processes we will below also use a procedure where we study a specific
number of initial-state splittings or only initial-state splittings
above a certain scale, $\mu_{exc}$, in which case we run the initial
state shower without modification to get the desired states and only
afterwards decide if the states should become exclusive.

This approach is efficient if the ratio $\frac{\sigma'^s}{\sigma'^i}$
does not heavily depend on the number of emissions.  This is normally
the case for higher emission multiplicities in contrast the first few
emissions where the interference effects play a role.  In the program
both approaches are implemented and we use each of them in such a
frequency so that the event weights variation is minimal.

\subsection{Modified luminosity for all possible helicity configurations}

Before proceeding to calculate $\sigma'^s$ in \eqref{eq:pexmod} we
have to generalise the expression to include all possible helicity
combinations contributing to the exclusive production.  First, we
define four kinds of the ``luminosity amplitudes'',
$\mathcal{L}^{(J_z)}$, for $J_z=\{0^+,0^-,+2^+,-2^+\}$, using the
notation in \eqref{eq:spinkin}:
\begin{eqnarray}
\mathcal{L}^{(J_z)} &=&
\int\frac{\dd^2 {\bf q} \quad2\spinkin_{J_z}({\bf q}_1,{\bf q}_2)}{{\bf q}^2 ({\bf q}-{\bf p}_1)^2 (-{\bf q}-{\bf p}_2)^2 } \\
&&\hspace{-1em}\phi_M (x_1,Q_1^2)  \phi_M (x_2,Q_2^2) 
\Tsud^{1/2}(Q_1^2,\mu_F^2)  \Tsud^{1/2}(Q_2^2,\mu_F^2) \nonumber
\end{eqnarray}

For the future use, it is more convenient to work with $\mathcal{L}$ directly related to some particular helicity configuration of the partons entering to the hard sub-process.
Let's define
\begin{eqnarray}
\mathcal{L}^{++} &=& \mathcal{L}^{(0^+)} + \mathcal{L}^{(0^-)}, \quad
\mathcal{L}^{--} = \mathcal{L}^{(0^+)} - \mathcal{L}^{(0^-)}, \\
\mathcal{L}^{-+} &=& \mathcal{L}^{(+2^+)},\qquad\qquad
\mathcal{L}^{+-} = \mathcal{L}^{(-2^+)}.
\end{eqnarray}
These relations allow us to rewrite the CEP cross section
(\ref{MasterExclusiveAmplitude}) as:
\begin{eqnarray}
  \sigma^{exc} &=& \int \dd y\, \dd\ln M^2\,  \dd^2{{\bf p}_1}  \dd^2{\bf p}_{2}\,
  \eu^{-b {\bf p}_1^2}  \eu^{-b {\bf p}_2^2}  \frac{1}{256\pi^2} \frac{1}{2M^2}\times \nonumber\\ 
   &\times&\sum_{\lambda, j}\Bigg|\sum_{a=1}^8
    \bigg(\mathcal{L}^{++} \Amp_{++\to \lambda}^{aa \to j} +
      \mathcal{L}^{+-} \Amp_{+-\to \lambda}^{aa \to j} + \nonumber\\
      &&\qquad\;+\mathcal{L}^{-+} \Amp_{-+\to \lambda}^{aa \to j} +
      \mathcal{L}^{--} \Amp_{-- \to \lambda}^{aa \to j} \bigg) \Bigg|^2
  \dd\Sigma_w ,
\end{eqnarray}
where $j$ and $\lambda$ denotes the colour state and helicity state of
all final state particles of the hard sub-process.  The index $i$
denotes the colour of the fusing gluons.

Finally, it is possible to formally factorise the cross
section formula into process independent luminosity part and the cross
section part:
\begin{eqnarray}
  \sigma^{exc} &=& \int \dd y\, \dd\ln M^2 \nonumber\\
  &&\int\frac{1}{\pi^2}\,  \dd^2{{\bf p}_1}  \dd^2{\bf p}_{2}\,
  \eu^{-b {\bf p}_1^2}  \eu^{-b {\bf p}_2^2}
  \sum_{\substack{\lambda_{l} \lambda_{r} \\ \lambda'_{l} \lambda'_{r} }} 
  \mathcal{L}^{\lambda_l\lambda_r} \mathcal{L}^{*\lambda'_l\lambda'_r} \nonumber\\
  &\times&\frac{1}{512M^2} 
  \sum_{\substack{\lambda, j\\a a'}}\Amp_{\lambda_l \lambda_r \to \lambda}^{aa \to j}
  \Amp_{\lambda'_l \lambda'_r \to \lambda}^{*a'a' \to j} \;\dd\Sigma_w \nonumber\\
  &\equiv& \int \dd y\, \dd\ln M^2\,\,  L^{exc}\!
  \left(\substack{\lambda_{l} \lambda_{r} \\ \lambda'_{l} \lambda'_{r} } \right)
  \,\sigma^{s}\!\left(\substack{\lambda_{l}\lambda_{r}\\ \lambda'_{l}\lambda'_{r} } \right)\,\dd\Sigma_w,
	\label{exclusive_noSelGeneral}
\end{eqnarray}
where there is an implicit summation in the last expression over all
four helicity indices, and the generalised colour singlet cross section
$\sigma^s$ incorporates the normalisation factor~$\frac{1}{512M^2}$.
The origin of this factor is explained below equation
(\ref{eq:spinSingletColSingletXsec}).
Notice, that the cross section formula (\ref{exclusive_noSelGeneral})
is a direct generalisation of relation (\ref{exclusive_noSel}), where
only the spin singlet component was considered.

A more general expression for the probability $p_{exc}$ of event
being exclusive is then
\begin{equation}
  p_{exc} = \frac{ L^{exc}\!\left(\substack{\lambda_{l}\lambda_{r}\\
        \lambda'_{l}\lambda'_{r}}\right)\,
    \sigma'^{s}\!\left(\substack{\lambda_{l}\lambda_{r}\\
        \lambda'_{l}\lambda'_{r}}\right)}{ L^{inc}\, \sigma'^i },
\end{equation}
where the modified inclusive cross section $\sigma'^i$ which
incorporates the splitting functions is defined by formula
(\ref{eq:sigmaMod}), whereas the corresponding cross section
$\sigma'^s$ will be derived in the next section.  Both luminosities
are evaluated at scale of the latest initial-state emission.

\section{Approximation of matrix elements using shower splittings}
\label{sec:matrix-elements}

To calculate the central exclusive cross section, the amplitude for
every helicity and colour combination of the studied sub-process must
be known, rather than the spin- and colour-averaged sub-process cross
section.  For $2\to 2$ processes the amplitude $A^{\lambda_{l1}
  \lambda_{r1} \to \lambda_3 \lambda_4}_{a_{l1} a_{r1} \to x_3 x_4 }$
depends on the helicities of the incoming ($\lambda_{l1},
\lambda_{r1}$) and outgoing particles ($\lambda_{3}, \lambda_{4}$) as
well as on the colours of the corresponding particles $a_{l1}$,
$a_{r1}$, $x_3$, $x_4$. The additional dependency on particle momenta
is not written out explicitly.

It is useful to introduce the ``generalised cross section'' $\sigma$ of
the hard sub-process:
\begin{eqnarray}
\sigma \left(\substack{a_{l1} a_{r1} \\ a'_{l1} a'_{r1}} \middle| \substack{\lambda_{l1} \lambda_{r1} \\ \lambda'_{l1} \lambda'_{r1}} \right) &=&\\
&&\hspace{-3em}\frac{1}{512M^2}\sum_{\substack{x_3 x_4\\ \lambda_3 \lambda_4}} A^{\lambda_{l1} \lambda_{r1} \to \lambda_3 \lambda_4}_{a_{l1} a_{r1} \to x_3 x_4 }\;\left( A^{\lambda'_{l1} \lambda'_{r1} \to \lambda_{3} \lambda_{4}}_{a'_{l1} a'_{r1} \to x_3 x_4} \right)^*\nonumber
\end{eqnarray}
which is summed over helicities and colours of the final state
particles but not over initial-state one.  Moreover, the colour and
helicity indexes of the incoming particles are in general considered
to be different for the amplitude and its complex conjugated.  The
generalisation of this cross section to $2 \to n$ processes is
straightforward.

Knowing the generalised cross section, the inclusive cross section
takes the form:
\begin{eqnarray}
  \sigma^{inc} =  \sum_{\substack{\lambda_{l1} \lambda_{r1} \\
      \lambda'_{l1} \lambda'_{r1} }}
  \delta_{\lambda_{l1} \lambda'_{l1}}  \delta_{\lambda_{r1} \lambda'_{r1}}
  \sum_{\substack{a_{l1} a_{r1} \\
      a'_{l1} a'_{r1}} } \delta_{a_{l1} a'_{l1} }
  \delta_{a_{r1} a'_{r1}} \nonumber \\
  \times\;\sigma\left(\substack{a_{l1} a_{r1} \\
      a'_{l1} a'_{r1} } \middle| \substack{\lambda_{l1} \lambda_{r1} \\
      \lambda'_{l1} \lambda'_{r1}} \right),
\end{eqnarray}
where the nominal and complex conjugate indexes were put to be equal
by means of delta functions.

The colour singlet spin singlet cross section $\sigma^{S(0^+)}$ is
obtained by an analogous formula (imposing ``left'' and ``right'' colours and
helicities to be identical):
\begin{eqnarray}
  \sigma^{S(0^+)} = \sum_{\substack{\lambda_{l1} \lambda_{r1} \\
      \lambda'_{l1} \lambda'_{r1} }}
  \delta_{\lambda_{l1} \lambda_{r1}}  \delta_{\lambda'_{l1} \lambda'_{r1}}
  \sum_{\substack{a_{l1} a_{r1} \\ 
      a'_{l1} a'_{r1}} } \delta_{a_{l1} a_{r1} } \delta_{a'_{l1} a'_{r1} } \nonumber\\
  \times\;\sigma\left(\substack{a_{l1} a_{r1} \\
      a'_{l1} a'_{r1}} \middle| \substack{\lambda_{l1} \lambda_{r1} \\
      \lambda'_{l1} \lambda'_{r1}} \right),
\end{eqnarray}
Or using another notation:
\begin{equation}
  \sigma^{S(0^+)} = 
  \sigma^{s}\left(\substack{++\\ ++ } \right) +
  \sigma^{s}\left(\substack{--\\ -- } \right) +
  \sigma^{s}\left(\substack{++\\ -- } \right) +
  \sigma^{s}\left(\substack{--\\ ++ } \right),
\end{equation}
where the summation over helicities is made explicit and
the generalised colour singlet cross section $\sigma^s$,
firstly used in (\ref{exclusive_noSelGeneral}), is:
\begin{equation}
  \sigma^{s} \left(\substack{\lambda_{l1} \lambda_{r1} \\ \lambda'_{l1} \lambda'_{r1} } \right) = 
   \sum_{\substack{a_{l1} a_{r1} \\
      a'_{l1} a'_{r1}} } \delta_{a_{l1} a_{r1} }
  \delta_{a'_{l1} a'_{r1}} \;\;  
  \sigma\left(\substack{a_{l1} a_{r1} \\
      a'_{l1} a'_{r1} } \middle| \substack{\lambda_{l1} \lambda_{r1} \\
      \lambda'_{l1} \lambda'_{r1}} \right).
\label{eq:GeneralisedColourSingletXsec}
\end{equation}

The aim of the following sections is to derive an approximative form
of the generalised cross section for the case where we have
initial-state parton shower splittings from the left and right
incoming partons. These emissions are assumed to be strongly ordered
in their $p_\perp$.

To be specific, let's consider the gluon emission from the ``left''
incoming parton. The new generalised cross section will then take the
form:
\begin{eqnarray}
  \sigma\left(\substack{a_{l2} a_{r1}
      \\ a'_{l2} a'_{r1}} \middle| \substack{\lambda_{l2} \lambda_{r1} \\
      \lambda'_{l2} \lambda'_{r1}} \right)
  &=& f_{a_{l2}, e_1, a_{l1}} f^*_{a'_{l2}, e_1, a'_{l1}} \\
  &&\hspace{-5em}P^{gg}_{\lambda^l_{e1}} \left( \substack{ z_{l1},\;\phi_{l1} \\
      \lambda_{l2} \to \lambda_{l1} } \right) 
  P^{gg*}_{\lambda^l_{e1}} \left( \substack{ z_{l1},\;\phi_{l1} \\ 
      \lambda'_{l2} \to \lambda'_{l1} } \right)\; 
  \hat{\sigma} \left(\substack{a_{l1} a_{r1} \\
      a'_{l1} a'_{r1}} \middle| \substack{\lambda_{l1} \lambda_{r1} \\
      \lambda'_{l1} \lambda'_{r1}} \right),\nonumber
\end{eqnarray}
where the splitting $g \to gg $ was considered.  Einstein summation
convention is employed, in particular it is summed over helicity
$\lambda^l_{e1}$ and colour $e_1$ of the emitted gluon.

For splittings that includes quarks, the SU(3) structure constants,
$f_{abc}$, must be replaced by the Gell-Mann matrices, $T^{a}_{ij}$, and
the splitting \emph{amplitude}, $P^{gg}$, by the corresponding
one.

It is obvious that the spin-momentum and colour parts describing
emissions factorise and the resulting generalised cross section after
$n-1$ emission from the left parton and $m-1$ emissions from the right
parton is given by:
\begin{eqnarray}
  \sigma\left(\substack{a_{ln} a_{rm} \\
      a'_{ln} a'_{rm} } \middle| \substack{\lambda_{ln} \lambda_{rm} \\ 
      \lambda'_{ln} \lambda'_{rm}} \right) &=&
  T_l^{em} \left( \substack{a_{ln} \to a_{l1} \\
      a'_{ln} \to a'_{l1} } \right)
  P_l^{em} \left( \substack{\lambda_{ln} \to \lambda_{l1}\\
      \lambda'_{ln} \to \lambda'_{l1}} \right)    \nonumber\\
  &\times& T_r^{em} \left( \substack{a_{rm} \to a_{r1} \\
      a'_{rm} \to a'_{r1}} \right)
  P_r^{em} \left( \substack{\lambda_{rm} \to \lambda_{r1}\\
      \lambda'_{rm} \to \lambda'_{r1}} \right) \nonumber\\
  &\times& \hat{\sigma}\left(\substack{a_{l1} a_{r1} \\
      a'_{l1} a'_{r1}} \middle| \substack{\lambda_{l1} \lambda_{r1} \\
      \lambda'_{l1} \lambda'_{r1}} \right),
\end{eqnarray}
where the colour emission tensor $T^{em}$ depends on type and order of
the splittings and its indexes correspond to nominal ($1 - 3$) or
adjoint ($1 - 8$) SU(3) representations depending on the type of the
first and last parton in the shower.  The space-like emission tensor
$P^{em}$ depends on type, order and kinematics of the splittings.
Both of these tensors will be discussed in the following.

\subsection{The colour emission tensor}
In case the first parton in the parton shower as well as the
parton entering to the hard sub-process are gluons, the general form
of the colour emission tensor can be expressed as:
\begin{eqnarray}
  T_{gg}^{em} \left(\substack{x_2 \to x_1\\x_4 \to x_3} \right) &=&
  A_{12} \Tr (T^{x_1} T^{x_2}) \Tr (T^{x_3} T^{x_4}) \\
  &+& A_{13} \Tr (T^{x_1} T^{x_3}) \Tr (T^{x_2} T^{x_4}) \nonumber\\
  &+& A_{14} \Tr(T^{x_1} T^{x_4}) \Tr(T^{x_2} T^{x_3}) \nonumber \\
  &&\hspace{-6.6em}+ B_{234} \Tr (T^{x_1} T^{x_2} T^{x_3} T^{x_4}) 
  +B_{243} \Tr(T^{x_1} T^{x_2} T^{x_4} T^{x_3}) \nonumber\\
  &&\hspace{-6.6em}+B_{324} \Tr(T^{x_1} T^{x_3} T^{x_2} T^{x_4}) 
  +B_{342} \Tr (T^{x_1} T^{x_3} T^{x_4} T^{x_2}) \nonumber\\
  &&\hspace{-6.6em}+B_{423} \Tr(T^{x_1} T^{x_4} T^{x_2} T^{x_3}) 
  +B_{432} \Tr (T^{x_1} T^{x_4} T^{x_3} T^{x_2}),\nonumber
\end{eqnarray}
where $\Tr (T^{x_1} T^{x_2}) = \frac{1}{2}\, \delta_{x_1 x_2}$. In our
notation the trace prescription is used for gluon colour indexes and
delta function for quark colour indexes that means especially that $\Tr (T^{x_1}
T^{x_1}) = 4$ and $\delta_{x_1 x_1} = 3$.  Consequently, any colour
emission tensor between gluon and gluon is fully determined by 9 real
numbers irrespectively on the parton shower composition between these
two gluons.  Especially, the colour emission tensor representing no
emissions in the parton shower has only one non-zero coefficient
$A_{12} = 4$.  A colour tensor with one extra gluon emission is related
to the original one by linear transformation (provided in 
\appref{sec:colo-emiss-tens}) applied to the coefficients $A_{ij}$, $B_{ijk}$.

In an analogous way, the colour emission tensor can be constructed
for the transition between (anti-)quark and (anti-)quark:
\begin{eqnarray}
  T_{qq}^{em} \left(\substack{i_2 \to i_1\\
      i'_2 \to i'_1} \right)
  &=& K_1\, \delta_{i_1 i'_1} \delta_{i_2 i'_2}+ \\
  &+& K_2\, \delta_{i_1 i'_2} \delta_{i_2 i'_1}
  + K_3\, \delta_{i_1 i_2} \delta_{i'_1 i'_2}. \nonumber
\end{eqnarray}
The case of no emissions will here corresponds to $K_3=1$, $K_{1,2}=0$
.

The last alternative is the transition between quarks and gluons which
is given by
\begin{eqnarray}
  T_{gq}^{em} \left(\substack{x_2 \to i_1\\
      x'_2 \to i'_1}\right)
  &=& D_{ij}\, \delta_{i_1 i'_1} \Tr T^{x_2} T^{x'_2} + \\ \nonumber
  &+& C_1\, ( T^{x_2} T^{x'_2} )_{i_1 i'_1} \;  +
  C_2\, ( T^{x'_2} T^{x_2} )_{i_1 i'_1} + \\ \nonumber
  &+&C_{1c}\, ( T^{x_2} T^{x'_2} )_{i'_1 i_1} +
  C_{2c}\, ( T^{x'_2} T^{x'_2} )_{i'_1 i_1} 
\end{eqnarray}
and
\begin{eqnarray}
  T_{qg}^{em} \left(\substack{i_2 \to x_1\\
      i'_2 \to x'_1}\right)
  &=& D_{ij}\, \delta_{i_2 i'_2}
  \Tr T^{x_1} T^{x'_1} + \\ \nonumber
  &+& C_1\, ( T^{x_1} T^{x'_1} )_{i_2 i'_2} \;+
  C_2\, ( T^{x'_1} T^{x_1} )_{i_2 i'_2} + \\ \nonumber
  &+& C_{1c}\, ( T^{x_1} T^{x'_1} )_{i'_2 i_2} +
  C_{2c}\, ( T^{x'_1} T^{x'_1} )_{i'_2 i_2},
\end{eqnarray}
where in the last expression the coefficients, $C_{1,2}$ are zero if
the first parton is a quark and $C_{1c,2c}$ are zero if the first parton
is an anti-quark.

In \appref{sec:colo-emiss-tens} we provide the complete list of linear
transformations, necessary for calculating the colour emission tensor
with an extra emission in the beginning of the shower.  Considering
the initial-state parton shower described by the tensor\footnote{The $p_f$ denotes flavour of parton which initiates the parton shower and $p_h$ is the flavour of parton entering to the hard sub-process. Both partons are assumed to be on the ``left'' or on the ``right'' side.} $T^{em}_{p_f p_h}$, if the first parton $p_f$ is
gluon it can be evolved backward to a quark, anti-quark, or a gluon,
making this parton the starting one.  A quark can be evolved backward
to a quark or a gluon and, finally, an anti-quark can be evolved to an
anti-quark or a gluon.  This gives in total 7~possibilities.
The parton $p_h$ attached to the hard sub-process can be quark,
anti-quark, or gluon making the overall
number of possible linear transformation equal to $7\times 3=21$.  In
reality, some of these transformations are independent of whether the
parton is quark or anti-quark which reduces the number of
non-identical linear transformations to 13.

In the procedure we have developed, the two colour emission tensors
are constructed, one for ``left'' side and one for ``right''.  Before
starting the backward parton shower evolution these tensors describe
the shower with zero emissions and the particular type ($T_{gg}^{em}$
or $T_{qq}^{em}$) is chosen according to the ``left'' and ``right'' type
of the parton entering to the hard sub-process.  After every step in
the backward evolution, one of these tensors is modified using the
appropriate transformation.

\subsection{The spin emission tensor}

The leading-order amplitudes corresponding to the possible splitting
will depend on the helicities of incoming, outgoing and emitted
parton, as well as on the momentum fraction $z$ and the azimuthal
angle $\phi$ of the emission. These can all be found in the
literature \cite{DelDuca:1999iql}.  Using these splitting amplitudes,
the spin emission tensor can be defined, and in the particular case of
only one emission, this tensor has a form:
\begin{equation}
  P^{1em} \left( \substack{\lambda_{2} \to \lambda_{1}\\
      \lambda'_{2} \to \lambda'_{1}} \right)
  = P_{\lambda_{e1}} \left( \substack{ z_{1},\;\phi_{1} \\
      \lambda_{2} \to \lambda_{1} } \right)
  P^*_{\lambda_{e1}} \left( \substack{ z_{1},\;\phi_{1} \\
      \lambda'_{2} \to \lambda'_{1} } \right),
\end{equation}
which depends on the helicity and ``conjugate'' helicity
of the incoming and outgoing parton.  If two emissions are considered
this tensor is determined by summing over the intermediate helicity
and ``conjugate'' helicity
\begin{equation}
  P^{2em} \left( \substack{\lambda_{3} \to \lambda_{1}\\
      \lambda'_{3} \to \lambda'_{1}} \right)
  = P^{1em} \left( \substack{\lambda_{3} \to \lambda_{2}\\
      \lambda'_{3} \to \lambda'_{2}} \right) 
  P^{1em} \left( \substack{\lambda_{2} \to \lambda_{1}\\
      \lambda'_{2} \to \lambda'_{1}} \right) 
\end{equation}

Introducing the helicity index $\bar{\lambda}$ which takes integer
values between 1 and 4 and incorporates information about helicity and
``conjugate'' helicity index the form of the last equation
\begin{equation}
  P^{2em}_{ \bar{\lambda}_{3}  \bar{\lambda}_{1} } =
  P^{1em}_{ \bar{\lambda}_{3}  \bar{\lambda}_{2} } \;
  P^{1em}_{ \bar{\lambda}_{2}  \bar{\lambda}_{1} } 
\end{equation}
resembles simple matrix multiplication.  To add the emission one
therefore only need to multiply the current spin tensor (matrix) with
matrix corresponding to the particular emission.  The different forms
of these matrices for all possible splittings provided in
\appref{sec:spin-emiss-tens}.

\subsection{Amplitude definition}
Within the process library the amplitude of each process is defined in
the colour trace basis, in a form similar to what used for example in
MadGraph\cite{Maltoni:2002qb}.  For the $gg \to gg$ process, which has
the most complicated colour topology, the colour trace basis has 6
terms.\footnote{The colour basis of, for example, $q\bar{q} \to gg$
  process has two terms only.}  It means that 6 amplitudes depending
on Mandelstam variables $s$, $t$, $u$ must be provided for every
helicity combination (together $6\times 16=96$ amplitudes).
Fortunately, most of these amplitudes are equal to zero or are
identical to each other, \eg\ due to parity invariance.

The general form of the amplitude decomposition in the colour trace
basis is
\begin{eqnarray}
  A^\lambda_{a_{l1} a_{r1} \to x_3 x_4}
  &=& A^\lambda_1 \, B^1_{a_{l1} a_{r1} \to x_3 x_4}
  + A^\lambda_2 \, B^2_{a_{l1} a_{r1} \to x_3 x_4} +\nonumber\\
  &&+ \cdots
  + A^\lambda_K \, B^n_{a_{l1} a_{r1} \to x_3 x_4},
\end{eqnarray}
where the $B^i$ are the colour basis vectors.  The amplitudes
$A^\lambda_i$ depend on Mandelstam variables and helicities of the
particles but not on their colours.

To calculate the generalised cross section, $\hat{\sigma}$, defined
above, the product of every two basis vectors $M^{ij}$ must be known:
\begin{equation}
  M^{ij} \left(\substack{a_{l1} a_{r1}\\
      a'_{l1} a'_{r1}} \right)
  = \sum_{x_3 x_4} B^i_{a_{l1} a_{r1} \to x_3 x_4} B^{*j}_{a'_{l1} a'_{r1} \to x_3 x_4}.
\end{equation}
This product does not depend on the final state colours but only on
the colours and ``conjugate'' colours of the incoming particles.  In
analogy with the procedure used for colour emission tensor, each of
these products can be expressed as a linear combination of few colour
tensors $\mathcal{B}_\alpha$ creating basis.

If the partons entering to the sub-process are gluons then the tensors
$\mathcal{B}_\alpha$ can be one of the following:
\begin{eqnarray}
  &&\Tr (T^{a_{l1}} T^{a_{r1}}) \Tr ( T^{a'_{l1}} T^{a'_{r1}} ) ,\;\;  \nonumber \\
  &&\Tr (T^{a_{l1}} T^{a'_{l1}}) \Tr ( T^{a_{r1}} T^{a'_{r1}} ),\;\;  \nonumber \\
  &&\Tr (T^{a_{l1}} T^{a'_{l1}}) \Tr ( T^{a_{r1}} T^{a'_{r1}} )  \nonumber \\
  &&\Tr(T^{a_{l1}} T^{a_{r1}}  T^{a'_{l1}}  T^{a'_{r1}}),\;\;   \nonumber \\
  &&\Tr(T^{a_{l1}}  T^{a_{r1}}  T^{a'_{r1}}  T^{a'_{l1}}),\;\;  \nonumber \\
  &&\Tr( T^{a_{l1}}  T^{a'_{l1}}  T^{a'_{r1}} T^{a_{r1}}),\;\;\mbox{or}   \nonumber \\
  &&\Tr(T^{a_{l1}} T^{a'_{r1}} T^{a'_{l1}} T^{a_{r1}}) \nonumber
\end{eqnarray}
and for sub-processes with the incoming quarks or anti-quarks the
possible colour tensors $\mathcal{B}_\alpha$ are:
\begin{equation}
  \delta_{a_{l1},a'_{l1}} \delta_{a_{r1},a'_{r1}}\, ,\;\,
  \delta_{a_{l1},a'_{r1}} \delta_{a_{r1},a'_{l1}}\, ,\;\,\mbox{or}\;\,
  \delta_{a_{l1},a_{r1}} \delta_{a'_{l1},a'_{r1}}.
\end{equation}
To be able to calculate the exclusive cross section, the amplitudes
$A^{\lambda_{l1} \lambda_{r1}\to \lambda_3 \lambda_4 }_i$ and the
linear decompositions of the $M^{ij}$ into colour tensor basis listed
above must be provided for each sub-process (see
\appref{sec:example-sub-process} for an example).  If the coefficients
in the decomposition of $M^{ij}$ are denoted as $M^{ij}_\alpha$ then
the generalised sub-process cross section, $\hat{\sigma}$, takes the
form:
\begin{eqnarray}
  \hat{\sigma} \left( \substack{a_{l1} a_{r1}\\
      a'_{l1} a'_{r1}} \middle|\substack{\lambda_{l1} \lambda_{r1} \\
      \lambda'_{l1} \lambda'_{r1} }  \right)
  = \frac{1}{512M^2}\sum_{ij,\,\alpha} M^{ij}_\alpha\,\mathcal{B}_\alpha\!\left(\substack{a_{l1} a_{r1}\\
      a'_{l1} a'_{r1}} \right)\times \nonumber \\
  \times \sum_{\lambda_3 \lambda_4}A_i^{\lambda_{l1}\lambda_{r1}\to\lambda_3\lambda_4 }\!(s,t,u)\;
  A_j^{*\lambda'_{l1} \lambda'_{r1}\to \lambda_3 \lambda_4 }\!(s,t,u)
\end{eqnarray}
and the generalised colour-singlet cross section, $\sigma^s$, defined in
(\ref{eq:GeneralisedColourSingletXsec}),
can be calculated (if initial-state radiation is present) using the
following formula:
\begin{eqnarray}
  \sigma^{s}\left(\substack{\lambda_{ln} \lambda_{rm} \\
      \lambda'_{ln} \lambda'_{rm} }\right)
  \;\;\;&=&\;\;\; P_l^{em} \left( \substack{\lambda_{ln} \to \lambda_{l1}\\
      \lambda'_{ln} \to \lambda'_{l1}} \right)    
  P_r^{em} \left( \substack{\lambda_{rm} \to \lambda_{r1}\\
      \lambda'_{rm} \to \lambda'_{r1}} \right) \nonumber\\
  &&\hspace{-7.5em}\times\;\delta_{a_{ln} a_{rm}} \delta_{a'_{ln} a'_{rm}}
   \times T_l^{em} \left( \substack{a_{ln} \to a_{l1}\\
      a'_{ln} \to a'_{l1}} \right)  
  \;\,T_r^{em} \left( \substack{a_{rm} \to a_{r1}\\
      a'_{rm} \to a'_{r1}} \right)  \nonumber\\
  &&\hspace{+0.0em}\times\;\hat{\sigma}\left(\substack{a_{l1} a_{r1} \\
      a'_{l1} a'_{r1}} \middle| \substack{\lambda_{l1} \lambda_{r1} \\
      \lambda'_{l1} \lambda'_{r1}} \right)
  \label{eq:masterGenCrossSec}
\end{eqnarray}
The term in the last line does not depend on the colours but only on
the helicities and ``conjugate'' helicities entering into the hard
sub-process.  To calculate this term, first the contractions between
colour emission tensors $T^{em}$ and all members of the
$\mathcal{B}_\alpha$ basis must be calculated.  Using these numbers
the coefficients $M^{ij}$ (the colour indexes were contracted) are
evaluated and consequently the whole term in the last line of
\eqref{eq:masterGenCrossSec}.  It represents 16 values corresponding
to all possible helicity combinations.  The full generalised colour
singlet cross section, $\sigma^s$, is finally obtained by multiplying
by the ``left'' and ``right'' spin emission matrices $P_l^{em}$ and
$P_r^{em}$.

\section{Sample results}
\label{sec:results}
In this section we present a few sample results from our
implementation of the Durham formalism.  We will focus the discussion
on the unique feature of our implementation, \ie\ possibility of
generation the exclusive states with higher particle multiplicities.
Currently, the process library includes all hard QCD $2\to 2$ processes;
Higgs boson production via $gg\to H$; the single $Z^0$ production via
$q\bar{q}\to Z^0$; and two photon production via $gg\to \gamma\gamma$
and $q\bar{q}\to \gamma\gamma$.  The program is modular, however, and
new processes can be easily added.

All presented calculations of the CEP cross sections are made
for $pp$ collisions at $\sqrt{s} = 13\,\TeV$ and are
based on MMHT2014 LO PDF \cite{Harland-Lang:2014zoa}.
They incorporate hadronisation as well as the final-state radiation.
The initial-state shower is evolved down to $\mu_{exc} = 1.5\,\GeV$
if not said otherwise.
All predictions incorporate the soft survival probability estimated
using veto on MPI.
This probability is around $0.06$ with little kinematic dependency.

\subsection{Di-jet production}
\label{sec:DijetProduction}

We start by studying the properties of our Monte Carlo model for
di-jet production at the LHC.  In contrast to other implementations of
the Durham formalism, our program allows for the generation the exclusive
di-jet event from any $2\to 2$ QCD hard sub-process, as long as
the partons which initiate the space-like parton shower are gluons
that can be in a colour singlet state.\footnote{There is a possible
  extension of this approach to showers initiated by $q\bar{q}$, where
  a screening quark rather than a screening gluon is exchanged in the
  loop to compensate the colour flow but this is not implemented in
  our current version.}

The variable which describes the size of the phase space available for
the space-like parton shower is  $\mu_{exc}$ which is the lowest
allowed transverse momentum of the emission.  The maximal allowed
$p_\perp$ of an emission is set to be equal to the hard scale of the
sub-process, given by the transverse momentum of the leading jet. For
ordinary inclusive events the cut-off scale for the initial-state
radiation (ISR) in \pytppp is
around $2\,\GeV$.  In our discussion, we study the events with
transverse momenta of the emissions starting at 1.5~GeV.

The inclusion of possible initial-state splittings in our approach
will naturally increase the exclusive cross section for di-jet
production, and cause a smearing towards low values in the
distribution of $M_{12}/M_X$.  The $M_{12}/M_{X}$ observable can be
seen as an experimental measure of the ``exclusivity'' of the
particular event.  $M_{12}$ is here the invariant mass of the two
leading jets, and the total mass, $M_X$, of the exclusive system $X$
can, in principle, be calculated from the outgoing protons relative
momentum loss, $\xi$, as $\sqrt{\xi_1 \xi_2 s}$.  Without any parton
showers this ratio equals to 1 on the parton-level.  The final-state
radiation and hadronisation can smear this distribution, especially if
the jet radius of the jet algorithm is small, since a final state
parton may radiate outside the jet cone, giving to the smaller value of
the invariant di-jet mass $M_{12}$.

\begin{figure*}
  \begin{minipage}{0.49\textwidth}
    \begin{center}
      \includegraphics[width=\linewidth]{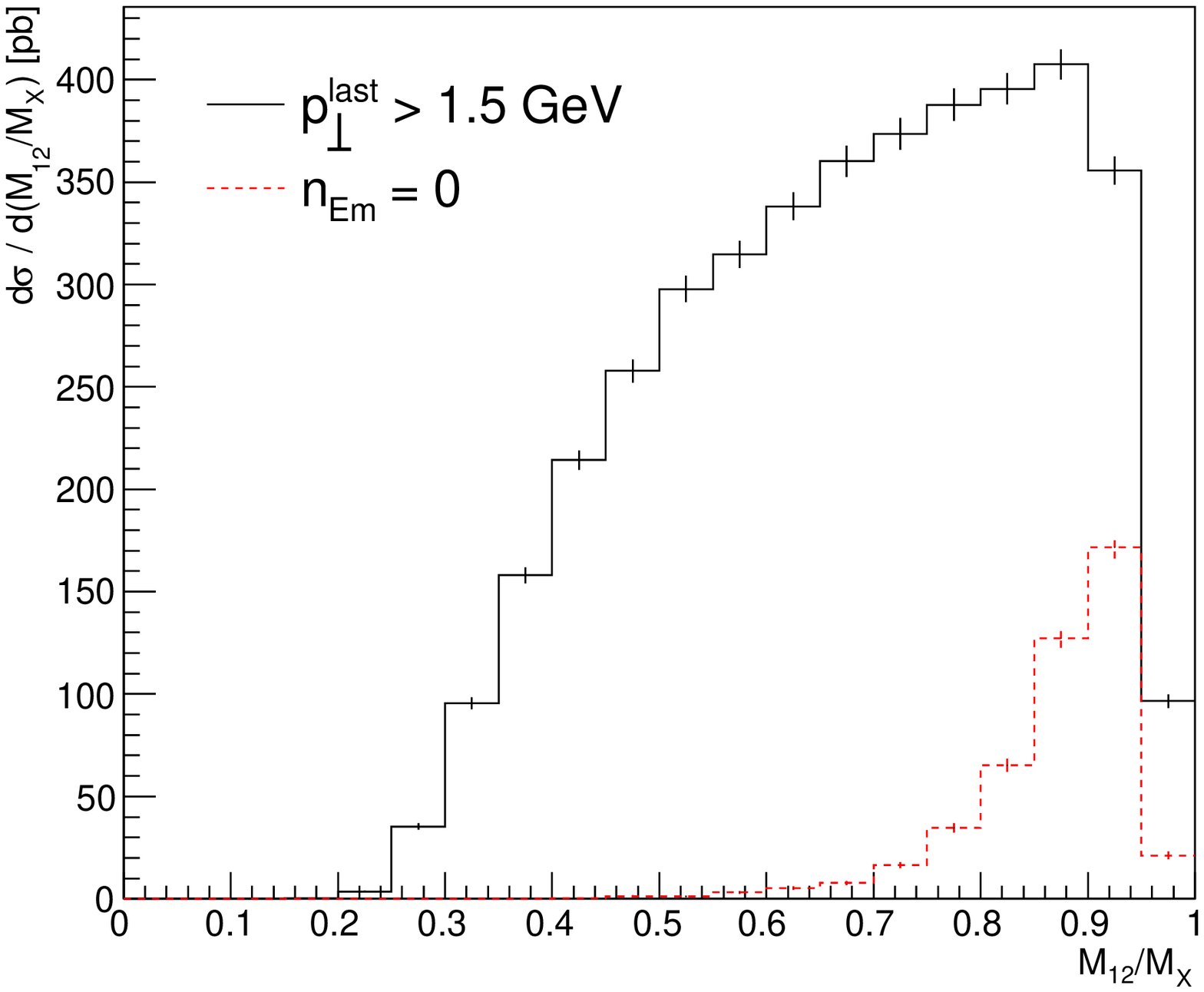}
    \end{center}
  \end{minipage}
  \begin{minipage}{0.49\textwidth}
    \begin{center}
      \includegraphics[width=\linewidth]{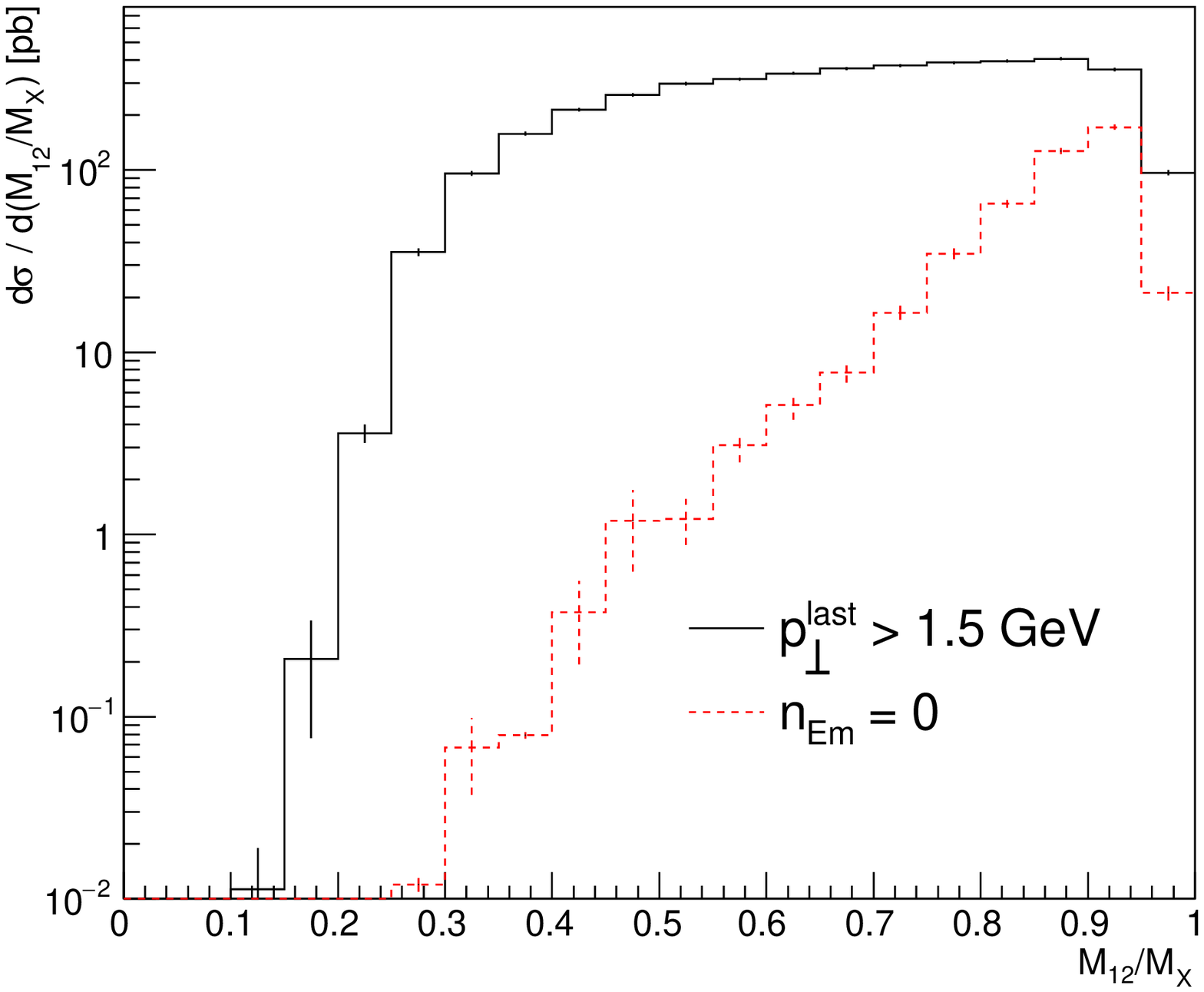}
    \end{center}
  \end{minipage}
  \caption{  The differential exclusive di-jet cross
    section for $pp$ collisions at $\sqrt{s}=13\,\TeV$ as
    a function of $M_{12}/M_X$. The phase space is defined by
    $p_\perp^{\mathrm{jet}1,2} > 40\,\GeV$,
    $|\eta^{\mathrm{jet}1,2}| < 2.5$ and $\xi_{1,2} < 0.03$. The blue
    curve represents the cross section with the final-state shower
    only in contrast to the black curve where the initial-state
    shower is included as well (down to $1.5\,\GeV$).
    Cross sections with no initial-state radiation correspond
    to the classical implementation of the Durham formalism.} 
  \label{fig:Mratio}
\end{figure*}

We have here used the ``anti-$k_\perp$'' jet algorithm
\cite{Cacciari:2008gp} with $R=0.7$, a minimum transverse momentum
of the jets of 40~GeV, and the absolute value of the pseudorapidity
of jets smaller than 2.5. As seen in \figref{fig:Mratio} there is indeed a
smearing from the final-state radiation (blue curve), but the smearing increases
significantly if initial-state radiation is included (black curve).
The distribution with initial-state radiation resembles what one would
expect form the double Pomeron scattering and the final states
generated by these two mechanisms overlap.
However, the physical nature of both processes are different since, in
DPE, where the Pomeron in the simplest approximation is a $gg$ object,
the colour neutralisation of the hard system comes from the Pomeron
remnant gluons, while for CEP it is due to an additional gluon
exchange. Despite different pictures, the final states could still be
indistinguishable, as low-$p_\perp$ initial state emission on either
side of the hard scattering in CEP could look exactly like Pomeron
remnants.

\begin{figure*}
  \begin{minipage}{0.49\textwidth}
    \begin{center}
      \includegraphics[width=\linewidth]{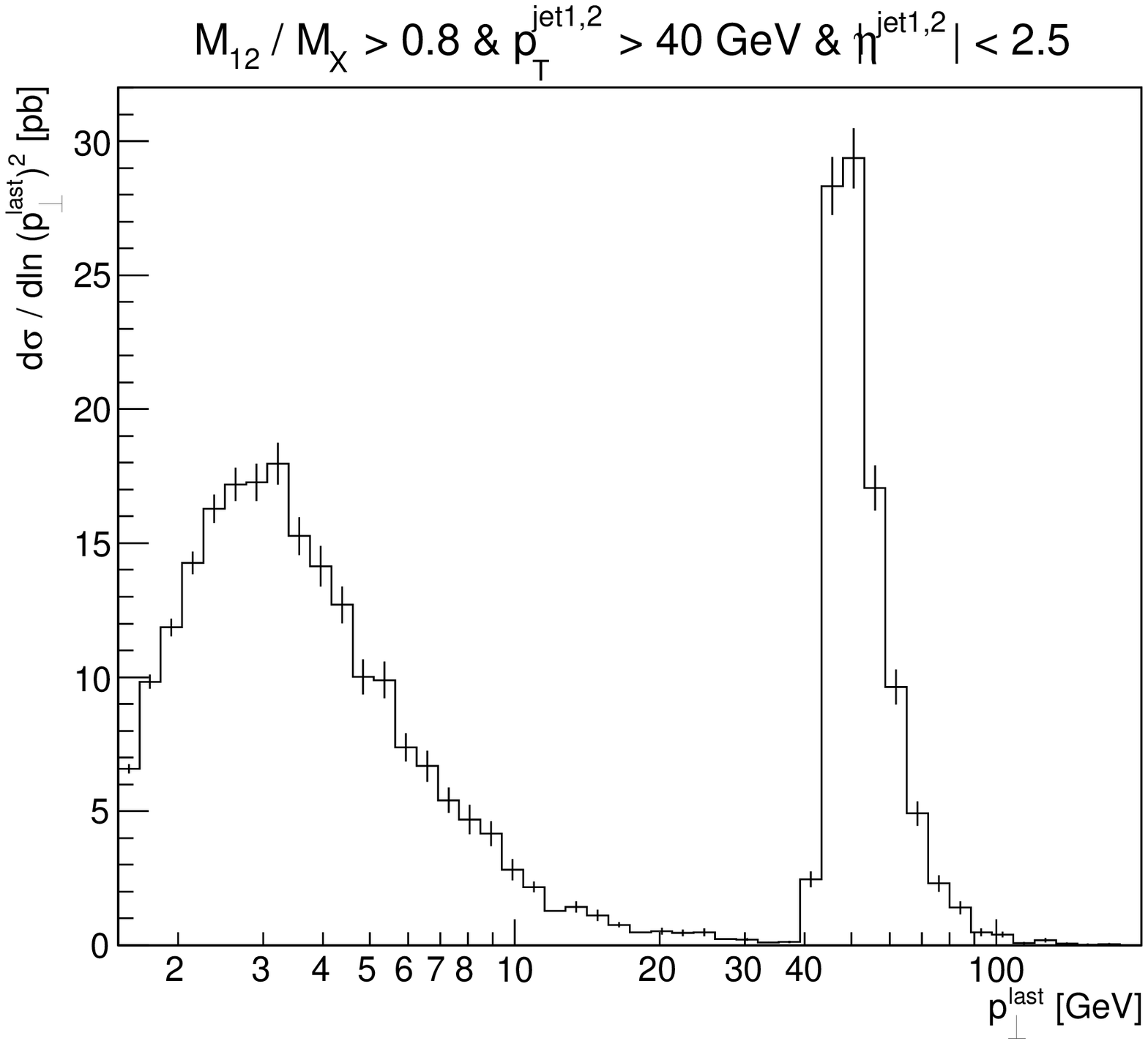}
    \end{center}
  \end{minipage}
  \begin{minipage}{0.49\textwidth}
    \begin{center}
      \includegraphics[width=\linewidth]{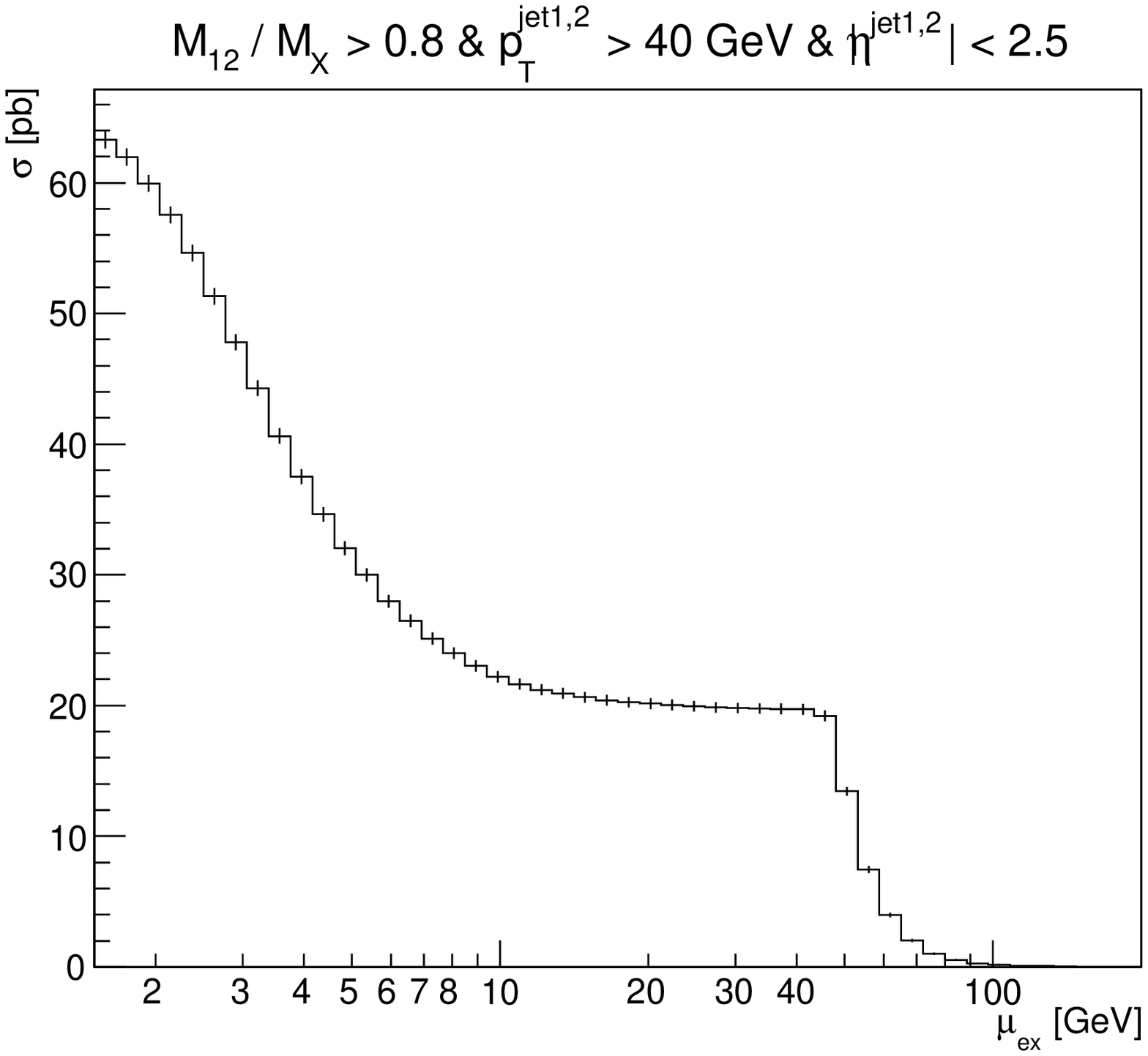}
    \end{center}
  \end{minipage}
  \caption{  The exclusive di-jet cross section
    for $pp$ collisions at $\sqrt{s}=13\,\TeV$ differential in
    $\ln (p_{\perp}^{\mathrm{last}})^2$ (left). The variable $p_\perp^{\mathrm{last}}$
		denotes the scale of the ``softest'' ISR emission.
		For events absent from
		any ISR emissions $p_\perp^{\mathrm{last}}$ is identified with the scale of
		the sub-process. 
		The total exclusive cross section as a
    function of the cut-off scale $\mu_{exc}$ is shown on the right.
		This scale means that that only events with
		$p_{\perp}^{\mathrm{last}} > \mu_{exc}$ are accepted.
    As is seen from the right plot the incorporation of the ISR 
    with $\mu_{exc} = 1.5\,\GeV$ increases
    the CEP cross section from $19\,\text{pb}$ to $63\,\text{pb}$.
		The phase
    space is defined by $p_\perp^{\mathrm{jet}1,2} > 40\,\GeV$,
    $|\eta^{\mathrm{jet}1,2}| < 2.5$, $M_{12}/M_X > 0.8$ and
    $\xi_{1,2} < 0.03$.  }
  \label{fig:SigmaExq2}
\end{figure*}

The exclusive cross section is less sensitive to the space-like
emissions if only the events with, for example $M_{12}/M_{X} > 0.8$
are accepted as is demonstrated in \figref{fig:SigmaExq2}.  

Here, the left plot shows that the di-jet cross section
consists of events either with $p_\perp^{\mathrm{last}} \sim 40\,\GeV$
where there was typically no space-like emission
and $p_\perp^{\mathrm{last}}$ was identified with the hard scale of the
process, or events with $p^{\mathrm{last}}_{\perp}\sim 3\,\GeV$.
In this case, there are usually many emissions and 
$p_\perp^{\mathrm{last}}$ denotes the transverse momentum of the latest one
with the smallest $p_\perp$.

It can be seen the di-jet cross section differential
in the $p_{\perp}$ of the last ISR emission peaks for
$p^{\mathrm{last}}_{\perp}\sim 3\,\GeV$
 and decreases for lower transverse momenta.

\begin{figure}
	\centering
	\includegraphics[width=\linewidth]{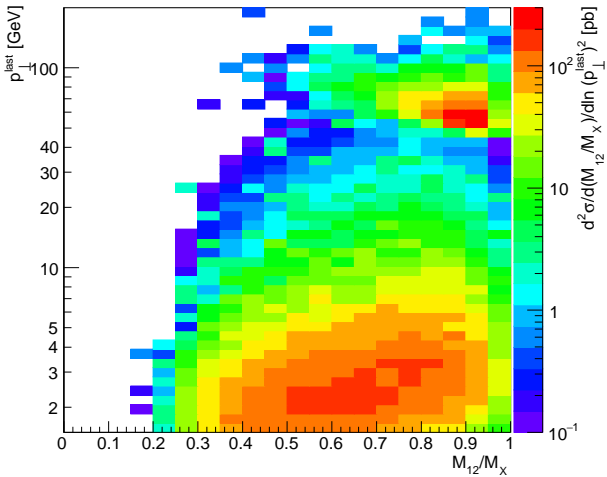}
  \caption{ The exclusive di-jet cross section for
    $pp$ collisions at $\sqrt{s}=13\,\TeV$ double differential
    in $M_{12}/M_X$ and $\ln\, (p_{\perp}^{\mathrm{last}})^2$.
		For sake of clarity the vertical axis is denominated
		directly in $p_\perp^\mathrm{last}$.
		The phase space is defined
    by $p_\perp^{\mathrm{jet}1,2} > 40\,\GeV$,
    $|\eta^{\mathrm{jet}1,2}| < 2.5$ and $\xi_{1,2} < 0.03$. }
   \label{fig:Mratio2d}
\end{figure}

More comprehensive picture of the situation provides the two-dimensional
plot (Fig.~\ref{fig:Mratio2d}), where the correlation of
the mass ratio and the $p_\perp^{\mathrm{last}}$ for a
particular event is shown.  The depletion of the emissions for
$p_\perp^\mathrm{last}$ slightly below $40\,\GeV$ is partially due to the
small $gg\to ggg$ colour singlet cross section and partially just a
statistical effects given by the low probability of having no further
emission below such high $p_\perp$.  The tree-level $gg\to
ggg$ spin-singlet colour-singlet cross section in the analytic form is
provided in \cite{Harland-Lang:2015faa}.  This cross section is zero
in the ``parton-shower'' limit where one of the outgoing gluons has
small $p_T$ compared to the remaining two and $\hat{t}=\hat{u}=-\hat{s}/2$,
where the Mandelstam variables are derived from the two hardest gluons
whereas the softest one is supposed to be part of the shower.  Such
behaviour agrees with our calculations based on procedure introduced
in section \ref{sec:matrix-elements}.

The cut-off parameter $\mu_{exc}$ can be understood as a variable which
describes the transition between perturbative and non-perturbative
region.
Not only due to the possible overlap with the double Pomeron exchange
process but also because the scale $\mu_{exc}$ denotes the minimal
allowed $p_\perp$ of the ISR emission and the $p_\perp$ of the
softest emission is simultaneously the highest allowed transverse
momentum of the screening gluon.
Choosing small $\mu_{exc}$ leads to low $p_\perp^\mathrm{last}$
and consequently the main contribution to the exclusive luminosity
given by integral (\ref{KMRintegrand}) stems from small
transverse momenta, \ie\ smaller than $1$~GeV,
 where the perturbative QCD is not justified \cite{Khoze:2004yb}.

Finally in table~\ref{tab:processes} we show the contribution to the
exclusive cross section from the different possible hard sub-process.

\begin{table*}
\centering
\caption{The table demonstrates how the particular
  hard sub-processes in \pytppp contribute to the total
  exclusive cross section of the di-jet production at LHC ($\sqrt{s} =
  13\,\TeV$). The hard processes are defined using the Pythia
  convention and are accompanied by the Pythia process Id \cite{PythiaQCD}. 
  The letter $q$ denotes any light quark flavour, therefore, \eg\ 
  $gg \to q\bar{q}$ represents the sum of $gg \to u\bar{u}$, $gg \to d\bar{d}$ and
  $gg \to s\bar{s}$ cross sections.  The jets
  in the di-jet system are required to have $p_\perp^{\mathrm{jet}1,2} >
  40\,\GeV$ and $|\eta^{\mathrm{jet}1,2}| < 2.5$. In addition
  the leading protons momentum loss $\xi$ must be $\xi_{1,2} <
  0.03$.
  The $\sigma^{n_{Em}=0}_{exc}$ are the exclusive cross
  sections with no initial-state radiation.	
  The $\sigma_{exc}$ and $\sigma^{M_{12}/M_X >0.8}_{exc}$ are the exclusive cross section with allowed
  initial-state radiation down to $1.5\,\GeV$; the last one 
	has an additional constrain $M_{12}/M_X > 0.8$. 
}
\label{tab:processes}
\begin{tabular}{|c|c|c|c|c|}
\hline
Id & Process          & $\sigma^{n_{Em}=0}_{exc}\; [\text{pb}]$ & $\sigma_{exc}\; [\text{pb}]$ &  $\sigma^{M_{12}/M_X >0.8}_{exc}\; [\text{pb}]$ \\
\hline
$111$ & $gg \to gg$               & $23$ & $173$ & $57$\\
$112$ & $gg \to q \bar{q}$        & $10.6\times 10^{-3}$ & $0.6$ & $56\times 10^{-3}$\\
$113$ & $qg \to qg$               & $-$ & $30$ & $5.8$\\
$114$ & $qq' \to qq'$             & $-$ & $1.3$ & $94\times 10^{-3}$\\
$115$ & $q\bar{q} \to gg$         & $-$ & $10.5\times 10^{-3}$ & $83\times 10^{-6}$\\
$116$ & $q\bar{q} \to q'\bar{q}'$ & $-$ & $16\times 10^{-3}$ & $0.5\times 10^{-3}$\\
$121$ & $gg \to c'\bar{c}'$       & $4.8\times 10^{-3}$ & $0.2$ & $21\times 10^{-3}$\\
$122$ & $q\bar{q} \to c'\bar{c}'$ & $-$ & $4.5\times 10^{-3}$ & $57\times 10^{-6}$\\
$123$ & $gg \to b'\bar{b}'$       & $20\times 10^{-3}$ & $0.3$ & $51\times 10^{-3}$\\
$124$ & $q\bar{q} \to b'\bar{b}'$ & $-$ & $4.4\times 10^{-3}$ & $53\times 10^{-6}$\\
\hline
& all & 23 & 205 & 63\\
\hline
\end{tabular}           
\end{table*}

One can see that even with space-like parton showers enabled, the
$gg\to gg$ sub-process dominates.  The second largest cross section is
given by the $qg\to qg$ process which is forbidden without ISR.
Consequently, the fraction of di-jet events where
at least one of them is quark-induced with respect to the total exclusive
di-jet cross section is much higher than $\sim\! 10^{-4}$ predicted
in~\cite{Harland-Lang:2014lxa}.
This fact makes it problematic to use the CEP as a pure source of gluonic 
jets.

Within the collinear approximation the $gg\to q\bar{q}$ cross section
is predicted to be suppressed as $m_q^2/s$ with respect to the
$gg\to gg$ cross section.
This is well-known consequence of the spin singlet selection rule.
It is interesting that without using such collinear approximation the
exclusive production of light flavour $q\bar{q}$ jets is not so heavily
suppressed since
the $|J_z|=2$ contribution, absent
in collinear case, has a similar size and is quark-mass independent
\cite{Harland-Lang:2014lxa}.
This effect is even stronger if the ISR is included.

Nevertheless, for higher $M_{12}/M_X$ the fraction of the heavy
flavours jets with respect to the whole CEP's di-jet sample
is still predicted to be lower compared to the DPE
which makes such quantity a vital experimental variable
for studying the transition region between CEP and DPE
as was first done at the Tevatron \cite{Aaltonen:2007hs}.

\begin{figure*}
  \begin{minipage}{0.50\textwidth}
    \begin{center}
      \includegraphics[width=\linewidth]{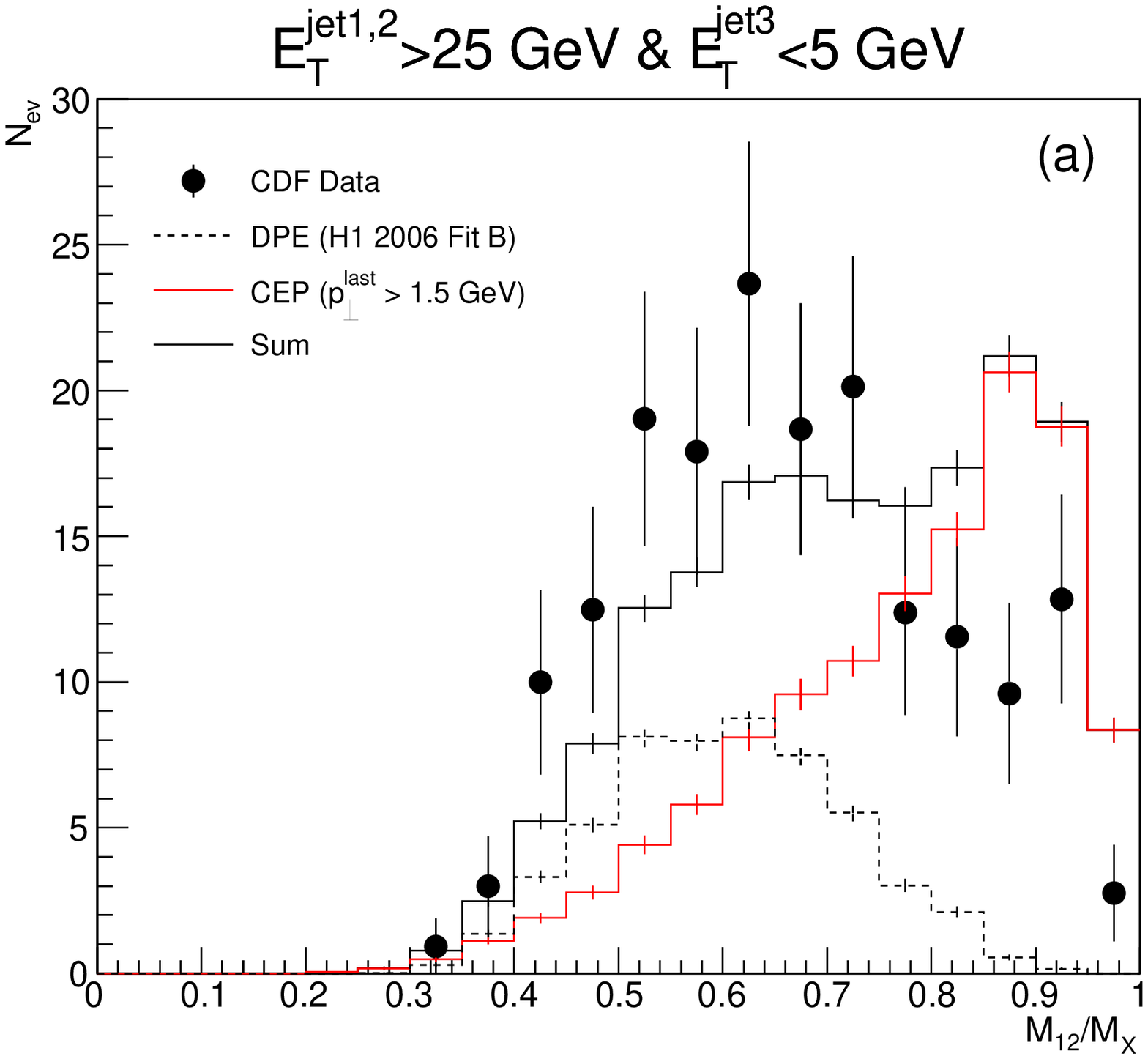}
    \end{center}
  \end{minipage}
  \begin{minipage}{0.50\textwidth}
    \begin{center}
      \includegraphics[width=\linewidth]{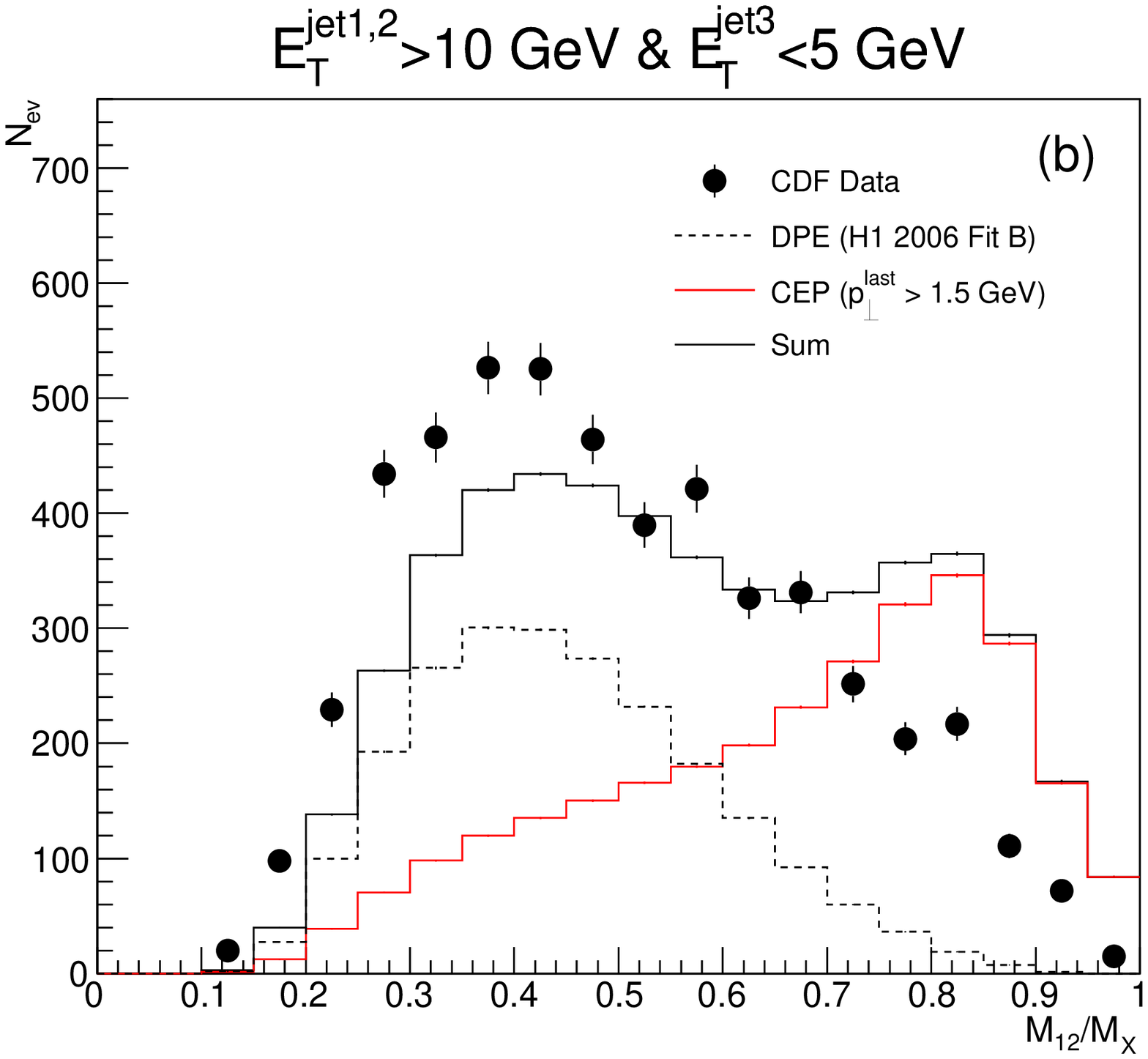}
    \end{center}
  \end{minipage}
  \begin{minipage}{0.50\textwidth}
    \begin{center}
      \includegraphics[width=\linewidth]{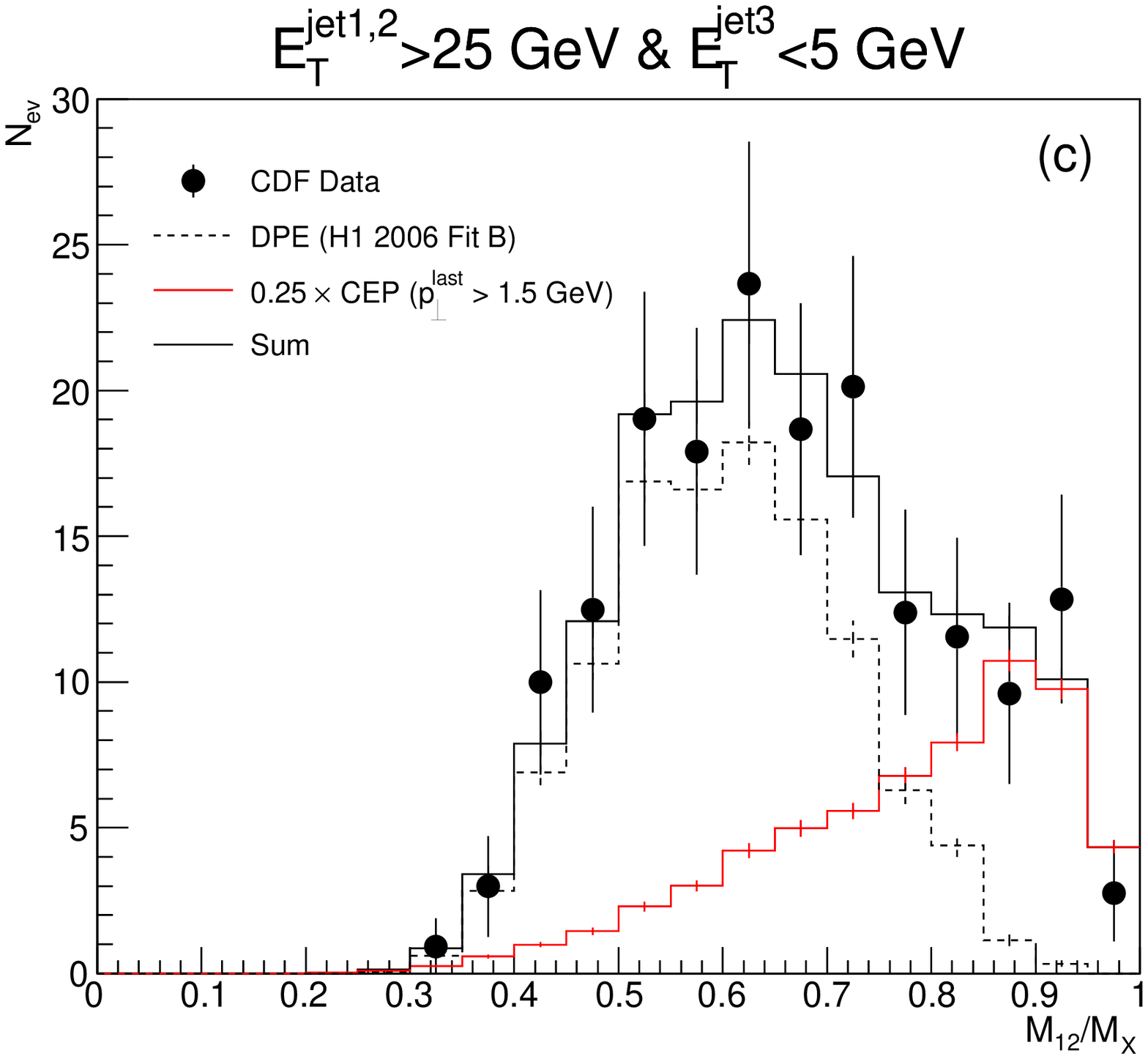}
    \end{center}
  \end{minipage}
  \begin{minipage}{0.50\textwidth}
    \begin{center}
      \includegraphics[width=\linewidth]{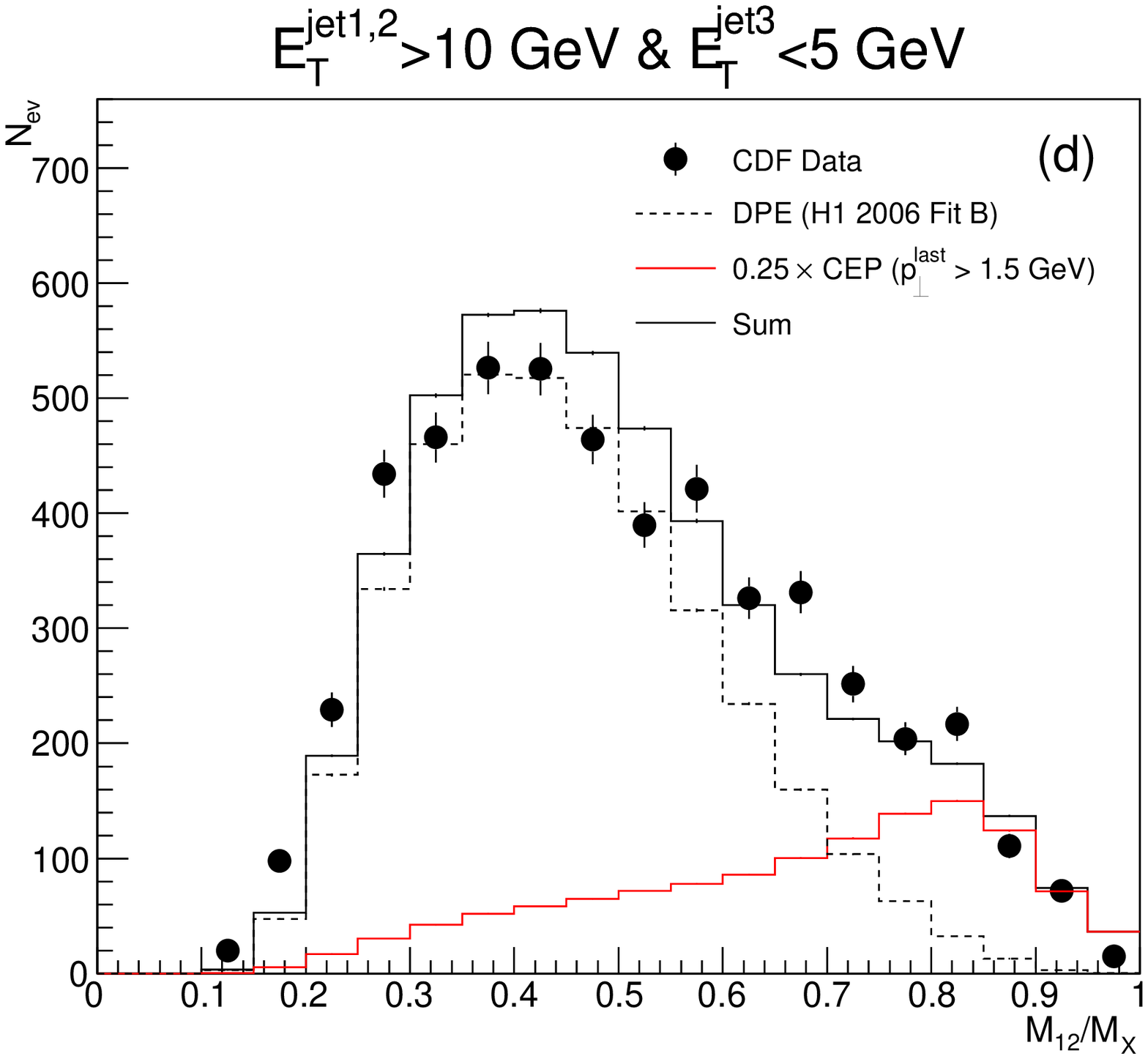}
    \end{center}
  \end{minipage}
  \caption{  The di-jet event counts binned in $M_{12}/M_X$ variable as 
		measured by the CDF collaboration \cite{Aaltonen:2007hs} in $p\bar{p}$ collisions at energy $\sqrt{s}=1.96\,\TeV$.
		These measured event counts (not corrected for the detector related effects) are compared with
		the DPE contribution (black dashed line) and CEP contribution (red line) and their sum (black solid line).
		The normalization of these curves is fixed in such a way that the total predicted event count is the same as in data.
		In figures c and d the CEP contributions were scaled down by a factor of $0.25$ compared to their nominal value.} 
   \label{fig:TevatronDijes}
\end{figure*}

In \figref{fig:TevatronDijes} we have tried to compare the results
from our CEP program for the distribution in $M_{12}/M_X$ with data
published by the CDF collaboration \cite{Aaltonen:2007hs}. The
comparison is a bit uncertain as the data has not been corrected to
the hadron-level, and the acceptance in different regions of phase
space is difficult to disentangle. Nevertheless we have checked that
our implementation of the DPE gives results similar to what was
published in \cite{Aaltonen:2007hs} for normalised distributions.

Our DPE implementation uses diffractive parton densities, as measured
\eg\ by HERA.  Specifically, we will here use the HERA H1 2006 Fit B
DPDFs \cite{Aktas:2006hy} with, for simplicity, the same soft survival
probability as for the CEP process, although we are aware that the
soft survival probability may very well be different for the DPE
process as compared to the CEP one. Technically the DPE simulation is
done in Pythia (with the same setting as for the CEP) by colliding two
hadrons, the Pomerons, with energies $\frac{1}{2}\xi_1 \sqrt{s}$ and
$\frac{1}{2}\xi_2 \sqrt{s}$ and with parton densities described by the
HERA DPDFs.\footnote{For single Pomeron processes this is now a
  standard option in \pytppp \cite{Rasmussen:2015qgr}, but we have
  here made our own simplified implementation of double Pomeron
  processes.}

In \figref{fig:TevatronDijes}a and b we show the results of simply
adding our CEP generated events for two different selection cuts (two
jets above $10$ and $25$~GeV respectively and no third jet above
$5$~GeV).  Further selection criteria, identical for both phase spaces
are given in \cite{Aaltonen:2007hs}.  We see that the addition of CEP
severely overshoots the data in the exclusive region of high
$M_{12}/M_X$. There are, however, many uncertainties, especially when
it comes to the soft survival probability, both for the CEP and DPE
contribution. As a demonstration we show in
\figref{fig:TevatronDijes}c and d the effect of introducing a relative
normalization factor of $0.25$ between the CEP and DPE contribution,
which gives a quite reasonable description of the data. We note that
our CEP, as expected, contributes quite noticeably also away from the
purely exclusive region. A more detailed study of the differences
between our new CEP procedure and the DPE one, especially in the
regions of the Pomeron remnants in DPE, may result in observables that
could further improve the experimental separation between the two
processes.

\subsection{Higgs production}

The possibility to measure the Higgs boson in the central exclusive
production was studied extensively in the last decade
\cite{Cox:2005if,Heinemeyer:2007tu}.  The discussion
was mainly focused on the dominant decay channel $gg\to H\to b
\bar{b}$ with a standard model branching ratio of $59\%$.

The main advantage of this production mechanism is a huge suppression
of the irreducible standard model background from $gg\to b\bar{b}$ due
to the $J_z=0$ selection rule in CEP.  Furthermore, the scalar nature
of the Higgs boson means that the ratio of exclusive to inclusive
cross sections is relatively enhanced as compared to the
background,\footnote{Quantitatively, $\frac{\hat{\sigma}^s}{\hat{\sigma}^i}(gg\to H) = 16$ (2 from spin $\times$ 8 from
  colour), whereas $\frac{\hat{\sigma}^s}{\hat{\sigma}^i}(gg\to b\bar{b}) = \frac{128}{7} \frac{m_b^2}{\hat{s}}\approx 0.02$.} as the spin and colours have to match also in the
inclusive sub-process. Both these effects improve the
signal/background ratio for the Higgs boson production compared to the
inclusive production.

The main background to the exclusive Higgs boson production comes from the
$gg\to gg$ sub-process, which can be substantially suppressed using
b-jet tagging techniques.  The other experimental challenge is the
detection of the scattered protons in the forward detectors in a high
pile-up environment where protons from several interactions can
simultaneously hit the forward detector within one bunch
crossing\footnote{The background protons typically originate from
  single diffractive excitation. Two such soft single diffractive
  events together with one inclusive can fake the CEP topology.
  Fortunately, this kind of experimental background is suppressed
  for higher masses of the exclusive system (higher $\xi$).}.

Note, that the cross section of the Higgs boson production in CEP is
only around $2\,\text{fb}$, including the calculated soft survival
probability of $0.06$, which is about four order lower than the
inclusive Higgs cross section $\sim 20\,\text{pb}$.
The signal event's count is further reduced due to selection criteria
and inefficiency of the b-jet tagging.  In particular, the QCD
background must be suppressed by selecting only high $p_\perp$ b-jets
(comparable to $M_H/2$) because the Higgs boson decay is isotropic
whereas the QCD jet production is suppressed at high $p_\perp$ at
least as $1./p_\perp^4$.

We included the Higgs boson production in the process library of our
program to study the production rates compared to the background
processes.  The simulation incorporates the parton showers as well as
hadronisation of the resulting partons into ``stable'' particles,
where the particles with lifetime higher than $0.01\,\text{mm/c}$ are
considered to be stable.  The inclusion of initial-state showers have
negligible effect on the exclusive Higgs cross section but can
substantially increase the $gg\to b\bar{b}$ background and spoil the
signal significance (see table~\ref{tab:processes}).

\begin{figure}
  \centering
  \includegraphics[width=\linewidth]{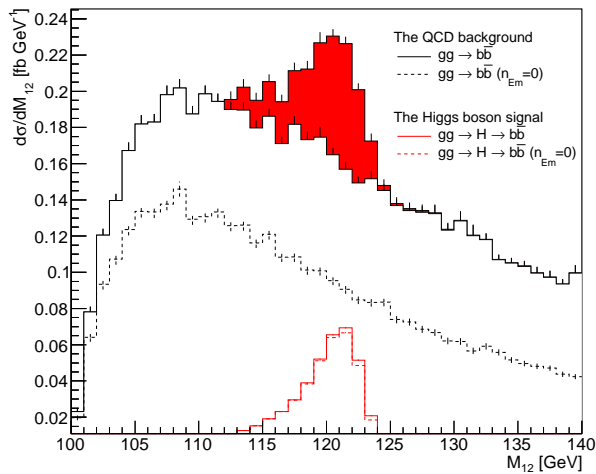}
  \caption{ The differential distribution
    of the invariant mass of two leading b-jets for $pp$ CEP at LHC
    ($\sqrt{s}=13\,\TeV$). To enhance the signal fraction the
    additional cuts $p^{\mathrm{jet}1,2}_{T}>50\,\GeV$ and $M_{12}/M_X >
    0.9$ were applied. The differential cross section stemming from
    the Higgs decay is given by the red solid curve and the QCD
    background from $gg\to b\bar{b}$ process is given by the black
    curve. In addition, the dashed curves indicate the corresponding
    cross section if initial-state showers are not considered.}
    \label{fig:HiggsProduction}
\end{figure}

We have simulated Higgs production at the LHC at $\sqrt{s}=13$~TeV.
The hard scale is set to be equal to the
Higgs mass for the signal, and to $p_\perp$ of the leading jet for the
background,  the cut-off for the space-like showers is 1.5~GeV in both
cases.  To pass the selection cuts, the events are required to contain
at least two b-jets with $p_\perp$ higher than 50~GeV and,
additionally, the ratio $M_{12}/M_{X}$ must be higher than 0.9.
As before, the jets are identified using anti-$k_\perp$ jet algorithm
with $R=0.7$ and a jet is tagged as a b-jet if it contains at least
one bottom hadron.  For
now, the kinematics of the scattered protons is not constrained.  The
result of these calculations is presented in
\figref{fig:HiggsProduction}, where the dotted lines indicate the
fraction of the cross section without space-like emissions.  It is
obvious that events with space-like emissions play a role only for the
$gg\to b\bar{b}$ process and their rate can be probably further
reduced using more sophisticated selection techniques.  Note that the
signal peak is a little bit shifted towards lower values compared to
the Higgs mass $m_{H} = 125$~GeV, due to the fact that sometimes not
all produced particles in the hadronisation of the b-quarks are
incorporated into the b-jets.

\begin{figure}
  \centering
  \includegraphics[width=\linewidth]{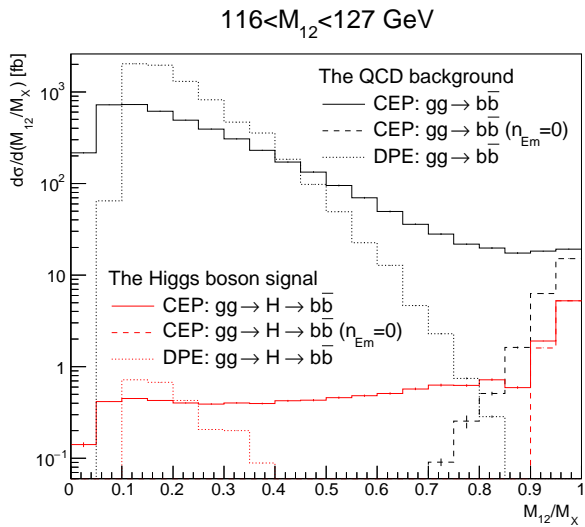}
  \caption{ The cross section for the CEP
    production of two b-jets at LHC with $p_\perp^{\mathrm{jet}1,2}>50$~GeV and
    invariant mass $116<M_{12}<127$~GeV differential in
    $M_{12}/M_X$. The red and black solid lines denote the cross
    section rising from the Higgs boson decay and the QCD background,
    respectively. The dashed lines denoted the corresponding cross
    sections with no initial-state radiation. In addition, the
    double Pomeron cross sections for Higgs production and the QCD
    $gg\to b\bar{b}$ process are plotted by the dotted lines. }
   \label{fig:HiggsPeak}
\end{figure}

In reality, the forward proton spectrometers installed to ATLAS and
CMS have a limited acceptance in $\xi$, the lowest measurable value of
$\xi$ is projected to be around $0.015$ which restricts the minimal
value of the exclusive system mass to $M_X=\xi \sqrt{s} =
195\,\GeV$.  This acceptance limit makes the observation of
single Higgs boson production without no other activity impossible.
On the other hand, there is still a hope of the signal of the Higgs
boson accompanied by jets originating from the space-like
emissions\footnote{Due to the colour singlet nature of Higgs
  production, at least two emissions are needed.}.  To see the size of
such cross section we plot the b-jets cross section (both of them must
still have $p_\perp>50\,\GeV$) in the mass window between $116$
and $127\,\GeV$ where the signal peak is expected.  This cross
section is shown in \figref{fig:HiggsPeak} as a function of the
$M_{12}/M_X$ ratio both for signal and background Monte Carlo sample.
The ratio of these cross sections roughly matches the
signal/background estimate. It is quite good for $M_{12}/M_X>0.9$
which is the kinematic phase space shown in
\figref{fig:HiggsProduction} whereas deteriorates for lower values of
$M_{12}/M_X$.

To reach the acceptance of LHC forward detectors, the mass ratio must
be lower than 0.6.  Assuming $0.5<M_{12}/M_X < 0.6$ the signal cross
section of the b-jets production is around 0.05~fb\footnote{Compare to
  0.4~fb for $M_{12}/M_X>0.9$.} and is around 200 times smaller than
the QCD background.  This small signal cross section and huge
background contamination leads to a luminosity of $\sim\!
60,000\,\text{fb}^{-1}$ to reach $4$-sigma precision.  Although the
possibility of measure the Higgs production in this experimental setup
is rather academic, our framework allows to determine such cross
sections as well as more realistically evaluate the contamination from
the QCD background processes.

In \figref{fig:HiggsPeak} we also show the corresponding calculation
from the DPE process, which becomes significant at low values of
$M_{12}/M_X$ both for the signal and background, but clearly does not
give any increase in the significance.

\subsection{{\boldmath $Z^0$} production}

Considering the acceptance of the forward proton spectrometers of the
ATLAS and CMS detectors, which is around $0.015<\xi<0.1$, the mass of
$Z^0$ resonance is much smaller than the acceptance limit $M_X\!\approx
\!200\,\GeV$.  This makes the study of direct production\footnote{Without
  additional hadronic activity.} of the $Z^0$ within the CEP mechanism
even more impossible than the Higgs.  Moreover, the $Z^0$ is
produced by $q\bar{q}\to Z^0$ sub-process,
which cannot be handled directly in the standard implementations of
the Durham model.

However, our model allows for initial-state radiation from the partons
entering to the hard sub-process, which can change the identity of
incoming quarks to gluons, which can be then treated using the
standard Durham exclusive luminosity. To do so, at least one $g\to
q\bar{q}$ emission from each side is needed.

Due to the colour singlet nature of the $Z^0$ and the fusing quarks,
there will probably be a non-negligible cross section for no space
like emissions and $q\bar{q}$ CEP luminosity with a screening quark as
discussed briefly above.  Here, we will make no attempt to evaluate
such cross section although our model can, in principle, be extended
to cover this production mechanism as well.

The other mechanism for central (semi-)exclusive production of a $Z^0$
is through DPE.  To estimate such a cross section we use the procedure
described in \sectref{sec:DijetProduction}, where we again we assume
that the DPE soft survival probability is the same as in CEP.
Contrary to the CEP where the $M_X$ mass is higher than $M_Z$ mostly
due to space like emissions, in DPE both the space like emissions and
the Pomeron remnants contribute to the mass.

We will look at semi-exclusive $Z^0\to\mu^{-}\mu^{+}$ production at
the LHC at $\sqrt{s}=13$~TeV, requiring a minimum transverse momentum
of $30$~GeV for the muons in the pseudorapidity region of
$|\eta|<2.5$. Both quasi-elastically scattered protons are required to
have $0.015<\xi_{1,2}<0.1$ in accordance to the acceptance of forward
proton spectrometers.  The electroweak process $q\bar{q}\to \mu
\bar{\mu}$ includes both $Z^0$ exchange and $\gamma$ exchange as well
as the interference terms.  We find that the resulting DPE cross
section is about ten times higher than for CEP. Quantitatively
the CEP cross section is around $3.5$~fb compared to $40$~fb for DPE.
The $Z^0$ can be produced also via the exclusive photoproduction.
The predicted cross section for this process, 
including $Z^0\to \mu^-\mu^+$ branching ratio, is, however,
about $0.3$~fb \cite{Motyka:2008ac} and would be even smaller if
the selection criteria for muons $p_T$ and pseudorapidity had been
applied.

\begin{figure*}[ht]
  \begin{minipage}{0.49\textwidth}
    \begin{center}
      \includegraphics[width=\linewidth]{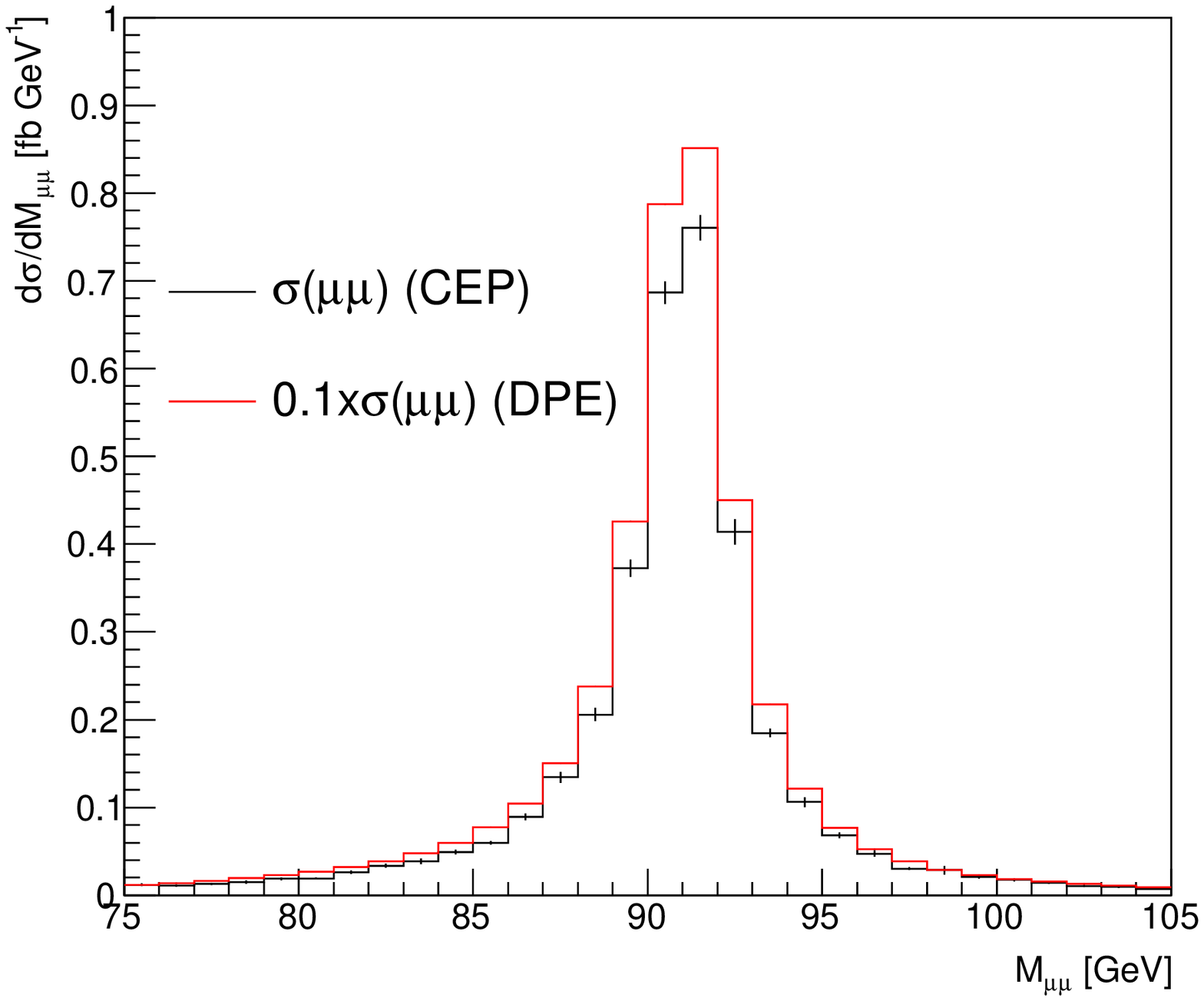}
    \end{center}
  \end{minipage}
  \begin{minipage}{0.49\textwidth}
    \begin{center}
      \includegraphics[width=\linewidth]{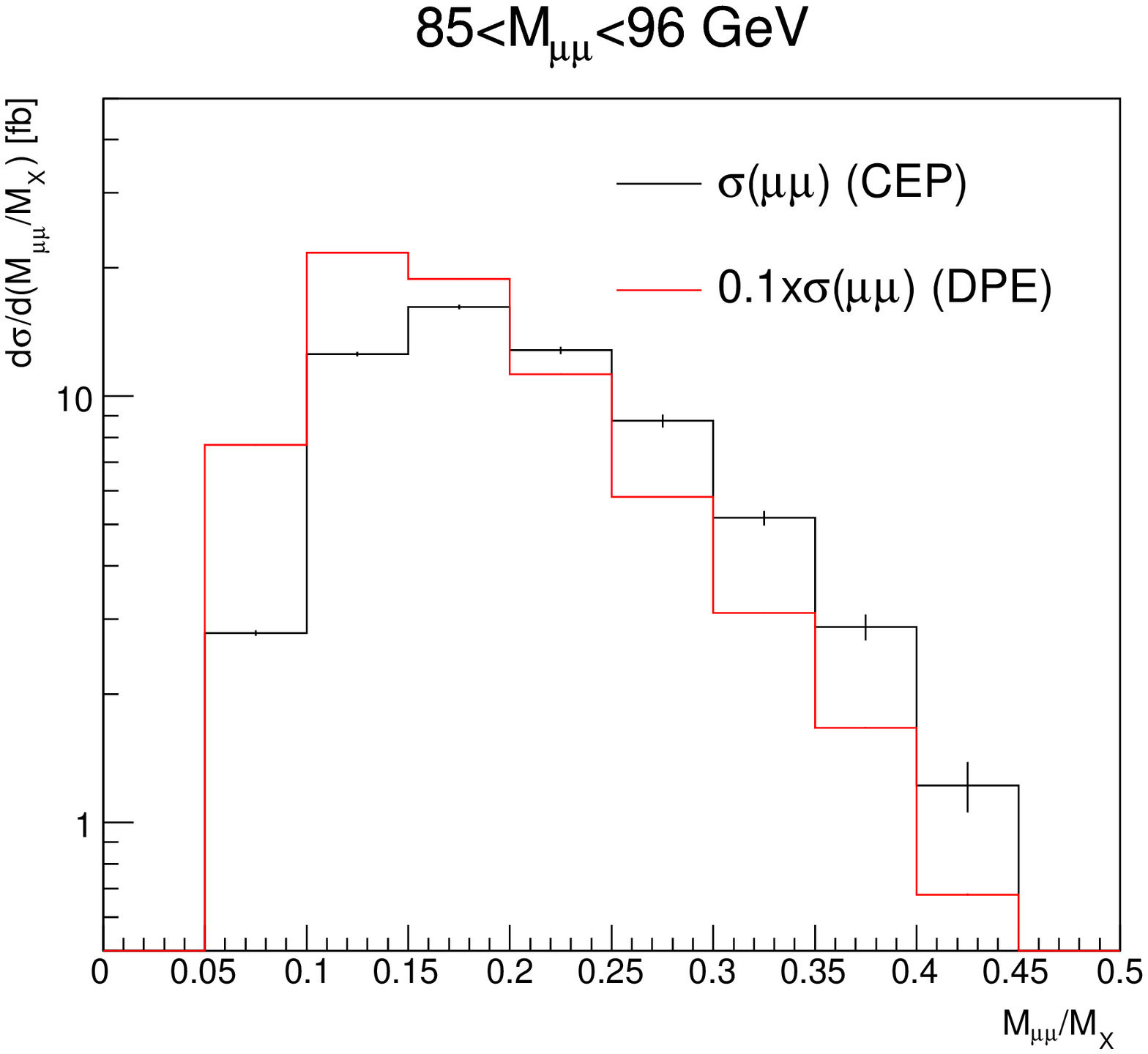}
    \end{center}
  \end{minipage}
  \caption{  The differential cross section of
    $pp\to p +\mu^{-}\mu^{+} X' +p$ process as a function of the
    invariant mass of the $\mu^{-}\mu^{+}$ pair (left plot) and of the
    ratio $M_{\mu\mu}/M_X$ (right plot). The black lines show the CEP
    contribution whereas the DPE result is given by the red lines. The
    DPE cross sections are normalised by a factor of $0.1$. In the CEP
    calculations the $\mu_{exc}$ cut-off parameter is set to 1.5~GeV.}
    \label{fig:Z0hists}
\end{figure*}

The shapes of the $M_{\mu\mu}$ and $M_{\mu\mu}/M_X$ distributions are
compared in \figref{fig:Z0hists}.  The shapes of $M_{\mu\mu}/M_X$ are
rather similar for both processes, with the DPE curve somewhat shifted
towards lower values as compared to the CEP one which prefers more
``exclusive'' configurations.

\section{Conclusions}
\label{sec:outlook}

In this paper we introduced a new Monte Carlo implementation of the
Durham formalism to calculate the central exclusive processes in $pp$
and $p\bar{p}$ collisions.  Our model is based on \pytppp generator,
and naturally incorporates partonic showers and hadronisation, as well
as multi-parton interactions.

The main advantage of our implementation is the possibility to study
the effects of initial-state parton radiation on CEPs. This is done by
allowing that any inclusively produced sub-process is converted to an
exclusive at any stage in the shower. To do this we have implemented a
colour and spin decomposition of the initial-state shower in \pytppp
which, together with a similarly decomposed (user supplied) matrix
element, can be used to determine the probability that a given
partonic state can be exclusive.

We have shown that this way of approximating higher jet
multiplicities gives rise to to new, non-trivial, physical consequences.  In
particular, for exclusive di-jet production, it leads to event
topologies with medium values of $M_{12}/M_X$ which naturally fill the
gap between double Pomeron exchange and pure central exclusive
production. Moreover, the incorporation of the parton showers enables
the generation of quark-initiated processes such as $Z^0$ production.

All predicted cross sections depend on the parameter $\mu_{exc}$, the
scale related the transition between the perturbative and
non-perturbative region in the parton shower.  The actual value of
$\mu_{exc}$ will have to be determined from experiment.  For the time
being we set its value equal to $1.5\,\GeV$.

The cross sections also depend on the soft survival probability
used. Here we have used the MPI model in \pytppp to simply estimate
the probability of having no additional scatterings, equating this to
the soft survival probability. Although this procedure was suggested
long ago, it has not been properly investigated, and we intend to
return with a detailed study of this model in a future publication.

Currently, the program process library includes QCD $2\to 2$
processes, $H$ production, $Z^0$ production and $\gamma\gamma$
production, but it can be easily extended. In particular, it would be
interesting to add production of vector mesons ($\rho$, $\phi$,
\ldots) and/or quarkonia $\chi_{c,b}$.  These processes have large
cross sections which make them experimentally accessible even at low
luminosities.

Our framework to treat colour and spin states within the partonic
shower is rather general and can, in principle, be extended to
simulate the central exclusive processes initiated by $q\bar{q}$
fusion, in addition to standard $gg$-initiated processes.  Here a
screening quark rather than screening gluon is exchanged to cancel the
colour flow. Such processes would be especially interesting for \eg\
central exclusive $Z^0$ production.

It should also be possible to extend our treatment of the colour and
spin structure of the parton showers to treat final-state
splittings. This would give an additional way of studying approximate
higher order effects in the hard sub-process matrix elements. 

These and other possible improvements will be discussed in a future
publication.

\begin{acknowledgements}
We are very grateful for many useful discussions with Tobj\"{o}rn
Sj\"{o}strand.  This work was supported in part by the MCnetITN FP7 Marie
Curie Initial Training Network, contract PITN-GA-2012-315877, the
Swedish Research Council (contracts 621-2012-2283 and 621-2013-4287).
\end{acknowledgements}

\appendix

\section{Colour emission tensors}
\label{sec:colo-emiss-tens}

In this section the list of all linear transformations which relate
the colour emission tensor before and after the emission is given.
New coefficients are labelled by the prime symbol.  If the expression for
any coefficient of the colour emission tensor is missing, this
coefficient is zero.

\begin{description}
\item[Adding of gluon:]\ 
\begin{itemize}
\item[$\Box$] $g \to (g \to g)$:
\begin{eqnarray}
A'_{12}  &=& - \frac{1}{2} B_{432} - \frac{1}{2} B_{234} \nonumber \\
A'_{13}  &=& + \frac{1}{2} B_{423} + \frac{1}{2} B_{342} + \frac{1}{2} B_{324} + \frac{1}{2} B_{243} + 3 A_{13}\nonumber \\
A'_{14}  &=& - \frac{1}{2} B_{432} - \frac{1}{2} B_{234}\nonumber  \\
B'_{234} &=& - \frac{1}{2} A_{14}  - \frac{1}{2} A_{12}\nonumber  \\
B'_{243} &=& + \frac{3}{2} B_{243} + \frac{1}{2} A_{12}\nonumber  \\
B'_{324} &=& + \frac{3}{2} B_{324} + \frac{1}{2} A_{14}\nonumber  \\
B'_{342} &=& + \frac{3}{2} B_{342} + \frac{1}{2} A_{12}\nonumber  \\
B'_{423} &=& + \frac{3}{2} B_{423} + \frac{1}{2} A_{14}\nonumber  \\
B'_{432} &=& - \frac{1}{2} A_{14}  - \frac{1}{2} A_{12}\nonumber
\end{eqnarray}
\item[$\Box$] $g\to (q \to q) \;\;\text{or}\;\; g\to (q \to \bar{q})$:
\begin{eqnarray}
    D'_{ij} &=& K_{1} \nonumber \\
    C'_1    &=& K_{3} \nonumber \\
    C'_{1c} &=& K_{2} \nonumber 
\end{eqnarray}
\item[$\Box$] $g\to (\bar{q} \to q) \;\;\text{or}\;\; g\to (\bar{q} \to \bar{q})$:
\begin{eqnarray}
    D'_{ij} &=& K_{1} \nonumber \\
    C'_{1c} &=& K_{3} \nonumber \\
    C'_{2}  &=& K_{2} \nonumber 
\end{eqnarray}
\item[$\Box$] $g\to (g \to q) \;\;\text{or}\;\; g\to (g \to \bar{q})$:
\begin{eqnarray}
    D_{ij} &=& \frac{1}{2} C_{2c} + \frac{1}{2} C_{1c} + \frac{1}{2} C_{2} + \frac{1}{2} C_{1} + 3 D_{ij} \nonumber \\
    C_{1}  &=& \frac{3}{2} C_{1} \nonumber \\
    C_{1c} &=& \frac{3}{2} C_{1c} \nonumber \\
    C_{2}  &=& \frac{3}{2} C_{2} \nonumber \\
    C_{2c} &=& \frac{3}{2} C_{2c} \nonumber
\end{eqnarray}
\item[$\Box$] $g\to (q \to g)$:
\begin{eqnarray}
    A'_{13}  &=& D_{ij} \nonumber \\
    B'_{243} &=& C_{2c} \nonumber \\
    B'_{324} &=& C_{1c} \nonumber
\end{eqnarray}
\item[$\Box$] $g\to (\bar{q} \to g)$:
\begin{eqnarray}
    A'_{13}  &=& D_{ij} \nonumber \\
    B'_{423} &=& C_{2} \nonumber \\
    B'_{342} &=& C_{1} \nonumber
\end{eqnarray}
\end{itemize}
\ 

\item[Adding of quark:]\
\begin{itemize}
\item[$\Box$] $q\to (q \to q)\;\;\text{or}\;\;q\to (q \to \bar{q})$:
\begin{eqnarray}
    K'_{1} &=& \frac{1}{2} K_{3} + \frac{1}{2} K_{2} + \frac{4}{3} K_{1} \nonumber \\
    K'_{3} &=& - \frac{1}{6} K_{3} \nonumber \\
    K'_{2} &=& - \frac{1}{6} K_{2} \nonumber
\end{eqnarray}

\item[$\Box$] $q\to (q \to g)$:
\begin{eqnarray}
    D'_{ij} &=& \frac{1}{2} C_{2c} + \frac{1}{2} C_{1c} + \frac{4}{3} D_{ij} \nonumber \\
    C'_{1c} &=& - \frac{1}{6} C_{1c} \nonumber \\
    C'_{2c} &=& - \frac{1}{6} C_{2c} \nonumber
\end{eqnarray}

\item[$\Box$] $q\to (g \to q)\;\;\text{or}\;\;q\to (g \to \bar{q})$:
\begin{eqnarray}
    K'_{1} &=&  \frac{5}{18} C_{2c} + \frac{1}{36} C_{1c} + \frac{5}{18} C_2 + \frac{1}{36} C_1 + \frac{2}{3} D_{ij} \nonumber \\
    K'_{3} &=& - \frac{1}{6} C_2  + \frac{7}{12} C_1 \nonumber \\
    K'_{2} &=& - \frac{1}{6} C_{2c} + \frac{7}{12} C_{1c} \nonumber
\end{eqnarray}

\item[$\Box$] $q\to (g \to g)$:
\begin{eqnarray}
    D'_{ij} &=& + \frac{1}{36} B_{432} + \frac{5}{18} B_{423} + \frac{5}{18} B_{342} +\nonumber\\&&+ \frac{1}{36} B_{324} + \frac{1}{36} B_{243} + \frac{1}{36} B_{234} + \frac{2}{3} A_{13} \nonumber \\
    C'_{1c} &=& - \frac{1}{12} B_{432} - \frac{1}{6} B_{342} + \frac{7}{12} B_{324} -\nonumber\\&&- \frac{1}{12} B_{234} + \frac{1}{4} A_{14} \nonumber \\
    C'_{2c} &=& - \frac{1}{12} B_{432} - \frac{1}{6} B_{423} + \frac{7}{12} B_{243} -\nonumber\\&&- \frac{1}{12} B_{234} + \frac{1}{4} A_{12} \nonumber
\end{eqnarray}
\end{itemize}
\ 

\item[Adding of anti-quark:]\ 
\begin{itemize}
\item[$\Box$]$\bar{q}\to (\bar{q} \to q)\;\;\text{or}\;\;\bar{q}\to (\bar{q} \to \bar{q})$:
\begin{eqnarray}
    K'_{1} &=& \frac{1}{2} K_{3} + \frac{1}{2} K_{2} + \frac{4}{3} K_{1} \nonumber \\
    K'_{3} &=& - \frac{1}{6} K_{3} \nonumber \\
    K'_{2} &=& - \frac{1}{6} K_{2} \nonumber
\end{eqnarray}

\item[$\Box$]$\bar{q}\to (\bar{q} \to g)$:
\begin{eqnarray}
    D'_{ij} &=& \frac{1}{2} C_{2} + \frac{1}{2} C_{1} + \frac{4}{3} D_{ij} \nonumber \\
    C'_{1} &=& - \frac{1}{6} C_{1} \nonumber \\
    C'_{2} &=& - \frac{1}{6} C_{2} \nonumber
\end{eqnarray}

\item[$\Box$]$\bar{q}\to (g \to q)\;\;\text{or}\;\;\bar{q}\to (g \to \bar{q})$:
\begin{eqnarray}
    K'_{1} &=& + \frac{1}{36} C_{2c} + \frac{5}{18} C_{1c} + \frac{1}{36} C_2 + \frac{5}{18} C_1 + \frac{2}{3} D_{ij} \nonumber \\
    K'_{3} &=& - \frac{7}{12} C_{2c}  - \frac{1}{6} C_{1c} \nonumber \\
    K'_{2} &=& - \frac{7}{12} C_{2}   - \frac{1}{6} C_{1} \nonumber
\end{eqnarray}

\item[$\Box$]$\bar{q}\to (g \to g)$:
\begin{eqnarray}
    D'_{ij}&=& + \frac{1}{36} B_{432} + \frac{1}{36} B_{423} + \frac{1}{36} B_{342} +\nonumber\\&&+ \frac{5}{18} B_{324} + \frac{5}{18} B_{243} + \frac{1}{36} B_{234} + \frac{2}{3} A_{13} \nonumber \\
    C'_{1} &=& - \frac{1}{12} B_{432} + \frac{7}{12} B_{342} - \frac{1}{6} B_{324} -\nonumber\\&&- \frac{1}{12} B_{234} + \frac{1}{4} A_{12} \nonumber \\
    C'_{2} &=& - \frac{1}{12} B_{432} + \frac{7}{12} B_{423} - \frac{1}{6} B_{243} -\nonumber\\&&- \frac{1}{12} B_{234} + \frac{1}{4} A_{14} \nonumber 
\end{eqnarray}
\end{itemize}
\end{description}

\section{Spin emission tensors}
\label{sec:spin-emiss-tens}

In this section the spin emission matrices are provided for all possible kinds of splittings.
In reality approximate the higher order matrix element squared, these splitting matrices should incorporate
an additional normalisation factor $(4\pi)^2 \frac{\alpha_s}{2\pi} \frac{1-z}{z} \frac{1}{p_\perp^2}$.
Since only the ratio $\frac{\sigma'^s}{\sigma'^i}$ is relevant in our framework
and the normalisation factors would be the same in numerator and denominator;
these factors can be simply omitted as they cancel in the ratio.

Note that the spin averaged splittings $P_{\mathrm{avg}}$ used in relation (\ref{eq:sigmaMod}) can be obtained (up to the normalisation arising from the colour part) as a sum of the ``corner'' elements of the spin emission matrix:
\begin{equation*}
P_{\mathrm{avg}} (z)  \sim  P^{1em}_{11} + P^{1em}_{14} + P^{1em}_{41}  + P^{1em}_{44} 
\end{equation*}

The spin emission matrices are the following:
\begin{equation*}
P^{1em}_{g\to g} =
\resizebox{20em}{!}{$
\left( \begin{matrix}
\frac{1}{z} + \frac{2}{1-z}  - 1  - z - z^2  &
-z (1-z)\, \eu^{ + 2 i \phi }  &
-z (1-z)\, \eu^{ - 2 i \phi }  &
\frac{1}{z} - 3 +  3 z - z^2
\\
- \frac{1-z}{z} \,  \eu^{- 2 i \phi} &
 \frac{2z}{1-z} &
 0 &
- \frac{1-z}{z} \,  \eu^{- 2 i \phi} 
\\
- \frac{1-z}{z} \,  \eu^{+ 2 i \phi} &
 0 &
 \frac{2z}{1-z} &
- \frac{1-z}{z} \,  \eu^{+ 2 i \phi} 
\\
\frac{1}{z} - 3 +  3 z - z^2 &
-z (1-z)\, \eu^{ + 2 i \phi }  &
-z (1-z)\, \eu^{ - 2 i \phi }  &
\frac{1}{z} + \frac{2}{1-z}  - 1  - z - z^2  
\end{matrix} \right)
$}
\end{equation*}

\begin{equation*}
P^{1em}_{g\to q} =
 \resizebox{18em}{!}{$
 \left( \begin{matrix}
   z^2 &
   -z (1-z) \eu^{ + 2 i \phi }  &
   -z (1-z) \eu^{ - 2 i \phi }  &
    (1-z)^2
   \\
   0 &
   0 &
   0 &
   0
   \\
   0 &
   0 &
   0 &
   0
   \\
   (1-z)^2 &
   -z(1-z)  \eu^{ + 2 i \phi } &
   -z(1-z)  \eu^{ - 2 i \phi } &
   z^2
\end{matrix} \right)
$}
\end{equation*}

\begin{equation*}
P^{1em}_{q\to g} = \left( \begin{matrix}
   \frac{1}{z} &
   0 &
   0 &
   \frac{(1-z)^2}{z} 
  \\
   -\frac{1-z}{z} \eu^{ - 2 i \phi } &
   0 &
   0 &
   -\frac{1-z}{z}  \eu^{ - 2 i \phi }
  \\
   -\frac{1-z}{z}  \eu^{ + 2 i \phi } &
   0 &
   0 &
   -\frac{1-z}{z}   \eu^{ + 2 i \phi }
  \\
   \frac{(1-z)^2}{z} &
   0 &
   0 &
   \frac{1}{z}
\end{matrix} \right)
\end{equation*}

\begin{equation*}
P^{1em}_{q\to q} = \left( \begin{matrix}
   \frac{1 + z^2}{1 - z} &
   0 &
   0 &
   0 
\\
   0 &
   \frac{2z}{1-z} &
   0 &
   0
\\
   0 &
   0 &
   \frac{2z}{1-z} &
   0
\\
   0 &
   0 &
   0 &
   \frac{1 + z^2}{1 - z} 
\end{matrix} \right)
\end{equation*}

\section{Example of the sub-process definition}
\label{sec:example-sub-process}

In this section an example of the sub-process definition is presented for $q \bar{q} \to g g$ process.
The amplitude of this process, written in the colour basis, has the following form:
\begin{equation*}
A^\lambda_{a_{l1}a_{r1}\to x_3 x_4} = A^\lambda_1\, (T^{x_3} T^{x_4} )_{a_{r1} a_{l1}} + A^\lambda_2\, (T^{x_4} T^{x_3} )_{a_{r1} a_{l1} },
\end{equation*}
where $\lambda$ denotes helicity state of both incoming and outgoing particles.
The indexes $a_{l1}$, $a_{r1}$ and $x_3$, $x_4$ denote the colour of incoming and outgoing particles
and $T^{x_{3,4}}$ are the Gell-Mann matrices.

Within our framework, the colour matrix $M^{ij}$ of the process must be provide by means of three colour basis vectors:
\begin{eqnarray}
\mathcal{B}_{ll'} &=& \delta_{a_{l1} a'_{l1}}  \delta_{a_{r1}  a'_{r1}}   \qquad \qquad
\mathcal{B}_{lr} = \delta_{a_{l1} a_{r1} }  \delta_{a'_{l1} a'_{r1} }  \nonumber \\
\mathcal{B}_{lr'} &=& \delta_{a_{l1} a'_{r1} }  \delta_{a'_{l1} a_{r1} }   \nonumber 
\end{eqnarray}
The colour matrix has the following form:
\begin{eqnarray}
\label{eq:ApCcolMatrix}
  M^{11} &=& \; \frac{7}{12} \mathcal{B}_{ll'} + \frac{1}{36} \mathcal{B}_{lr}  \quad\, 
  M^{12} = - \frac{1}{6} \mathcal{B}_{ll'} + \frac{5}{18} \mathcal{B}_{lr}  \\
  M^{21} &=& - \frac{1}{6} \mathcal{B}_{ll'} + \frac{5}{18} \mathcal{B}_{lr} \quad\,
  M^{22} =  \frac{7}{12} \mathcal{B}_{ll'} + \frac{1}{36} \mathcal{B}_{lr}\nonumber
\end{eqnarray}
Therefore, for example, the $M^{11}_\alpha$ coefficients for this process are:
\begin{equation}
M^{11}_{ll'} = \frac{7}{12} \qquad  M^{11}_{lr} = \frac{1}{36} \qquad M^{11}_{lr'} = 0.
\end{equation}

In addition to the colour matrix, the amplitudes for every helicity configuration must be given as well.
These amplitudes are:
\begin{eqnarray}
A_1^{+-\to +-}   &=&+\frac{1}{\sqrt{2}}\, 2 g^2  \frac{1}{s} \sqrt{tu}\, \eu^{i\phi} \label{eq:ApCamplitudes1a} \\
A_2^{+-\to +-}   &=&-\frac{1}{\sqrt{2}}\, 2 g^2  \frac{u}{s} \sqrt{\frac{u}{t}}\,\eu^{i\phi} 
\label{eq:ApCamplitudes1b}
\end{eqnarray}
\begin{eqnarray}
A_1^{+-\to -+}   &=&+\frac{1}{\sqrt{2}}\, 2 g^2  \frac{t}{s} \sqrt{\frac{t}{u}}\,\eu^{i\phi}\label{eq:ApCamplitudes2a} \\
A_2^{+-\to -+}   &=&-\frac{1}{\sqrt{2}}\, 2 g^2  \frac{1}{s} \sqrt{tu}\,\eu^{i\phi} 
\label{eq:ApCamplitudes2b}
\end{eqnarray}
The $s$, $t$ and $u$ are the Mandelstam variables, $\phi$ is the azimuthal angle of first
outgoing gluon and $g=\sqrt{4\pi \alpha_s}$.
The amplitudes related by a parity transformation $-+\to -+$ and $-+\to +-$ can be obtained by the complex conjugation. The amplitudes of other helicity configurations are equal to zero.
The normalisation factor $\frac{1}{\sqrt{2}}$ accounts for the identical particles in the final state.
The fact that the outgoing particles are identical allows to derive the amplitudes (\ref{eq:ApCamplitudes2a}-\ref{eq:ApCamplitudes2b}) for the second helicity configuration $+-\to -+$ from the first one, eq. (\ref{eq:ApCamplitudes1a}-\ref{eq:ApCamplitudes1b}), by swapping the final state gluons ($t\leftrightarrow u$).

To summarise, each process in the process library is defined by the helicity amplitudes (\ref{eq:ApCamplitudes1a}-\ref{eq:ApCamplitudes2b}) and the corresponding colour matrix (\ref{eq:ApCcolMatrix}).

\bibliographystyle{spphys}       
\bibliography{refs}   

\end{document}